\documentclass{aa}  

\usepackage{graphicx}
\usepackage{txfonts}
\usepackage{natbib}

\usepackage{tabularx}
\usepackage{wasysym}

\usepackage[colorlinks, citecolor=blue, linkcolor=blue]{hyperref}
\hypersetup{colorlinks,breaklinks, linkcolor=blue,urlcolor=magenta,,citecolor=blue}

\def\gcm3{\hbox{g cm$^{-3}$}}       
\def\Msun{\hbox{$\mathrm{M}_{\astrosun}$}}             
\def\Rsun{\hbox{$\mathrm{R}_{\astrosun}$}}
\def\Mjup{\hbox{$\mathrm{M}_{\rm J}$}}
\def\Rjup{\hbox{$\mathrm{R}_{\rm J}$}}

\bibpunct{(}{)}{;}{a}{}{,} 

\newcommand{\be}{\begin{equation}}
\newcommand{\ee}{\end{equation}}

\newcommand{\koi}{KOI-3886}
\newcommand{\kA}{KOI-3886\,A}
\newcommand{\kB}{KOI-3886\,B}
\newcommand{\kC}{KOI-3886\,C}

\begin{document} 


   \title{Uncovering the ultimate planet impostor\thanks{Based on observations collected at Centro Astron\'omico Hispano en Andaluc\'ia (CAHA) at Calar Alto, operated jointly by Instituto de Astrof\'isica de Andaluc\'ia (CSIC) and Junta de Andaluc\'ia and observations made with the Mercator Telescope, operated on the island of La Palma by the Flemish Community, at the Spanish Observatorio del Roque de los Muchachos of the Instituto de Astrofísica de Canarias.}}
   \subtitle{An eclipsing brown dwarf in a hierarchical triple with two evolved stars}

   \author{
   J.~Lillo-Box\inst{\ref{cab}}, 
   \'A.~Ribas\inst{\ref{eso}}, 
   B.~Montesinos\inst{\ref{cab}}, 
   N.~C.~Santos\inst{\ref{porto},\ref{depfisporto}}, 
   T.~Campante\inst{\ref{porto},\ref{depfisporto}}, 
   M.~Cunha\inst{\ref{porto}}, 
   D.~Barrado\inst{\ref{cab}}, \\ 
   E.~Villaver\inst{\ref{cab}}, 
   S.~Sousa\inst{\ref{porto}}, 
   H.~Bouy\inst{\ref{bordeaux}}, 
   A.~Aller\inst{\ref{cab},\ref{svo}}, 
   E. Corsaro\inst{\ref{inaf}},
   T. Li\inst{\ref{bham},\ref{aarhus}},
   J.~M.~J.~Ong\inst{\ref{yale}},\\
   I.~Rebollido\inst{\ref{stsci}}, 
   J.~Audenaert\inst{\ref{leuven}}, 
   F.~Pereira\inst{\ref{porto}} 
}

\institute{
Centro de Astrobiolog\'ia (CAB, CSIC-INTA), Depto. de Astrof\'isica, ESAC campus 28692 Villanueva de la Ca\~nada (Madrid), Spain\label{cab} \email{Jorge.Lillo@cab.inta-csic.es  }
\and European Southern Observatory, Alonso de Cordova 3107, Vitacura, Region Metropolitana, Chile \label{eso}
\and Instituto de Astrof\' isica e Ci\^encias do Espa\c{c}o, Universidade do Porto, CAUP, Rua das Estrelas, PT4150-762 Porto, Portugal \label{porto} 
\and Depto. de F\'isica e Astronomia, Faculdade de Ci\^encias, Universidade do Porto, Rua do Campo Alegre, 4169-007 Porto, Portugal \label{depfisporto}
\and Laboratoire d’astrophysique de Bordeaux, Univ. Bordeaux, CNRS, B18N, allée Geoffroy Saint-Hilaire, 33615 Pessac, France \label{bordeaux}
\and Spanish Virtual Observatory, Spain \label{svo} 
\and INAF -- Osservatorio Astrofisico di Catania, Via S. Sofia 78, 95123 Catania, Italy \label{inaf} 
\and School of Physics and Astronomy, University of Birmingham, Edgbaston, B15 2TT, UK \label{bham} 
\and Stellar Astrophysics Centre, Department of Physics and Astronomy, Aarhus University, 8000 Aarhus C, Denmark \label{aarhus}
\and Department of Astronomy, Yale University, 52 Hillhouse Ave., New Haven, CT 06511, USA \label{yale}
\and Space Telescope Science Institute, 3700 San Martin Drive, Baltimore, MD 21218, USA \label{stsci}
\and Institute of Astronomy, KU Leuven, Celestijnenlaan 200D, BUS-2410, Belgium \label{leuven} 
}

   \date{in prep}

 
  \abstract
   {Exoplanet searches through space-based photometric time series have shown to be very efficient in recent years. However, follow-up efforts on the detected planet candidates have been demonstrated to be critical to uncover the true nature of the transiting objects.}
   {In this paper we show a detailed analysis of one of those false positives hidden as planetary signals. In this case, the candidate KOI-3886.01 showed clear evidence of a planetary nature from various techniques. Indeed, the properties of the fake planet set it among the most interesting and promising for the study of planetary evolution as the star leaves the main sequence.}
   {To unveil the true nature of this system, we present a complete set of observational techniques including high-spatial resolution imaging, high-precision photometric time series (showing eclipses, phase curve variations, and asteroseismology signals), high-resolution spectroscopy, and derived radial velocities to unveil the true nature of this planet candidate. }
   {We find that KOI-3886.01 is an interesting false positive case: a hierarchical triple system composed by a $\sim$K2~III giant star (\kA{}) accompanied by a close-in eclipsing binary formed by a subgiant $\sim$G4~IV star (\kB{}) and a brown dwarf (\kC{}). In particular, \kC{} is one of the most irradiated brown dwarfs known to date, showing the largest radius in this substellar regime. It is also the first eclipsing brown dwarf known around an evolved star.}
   {In this paper we highlight the relevance of complete sets of follow-up observations to extrasolar planets detected by the transit technique using large-pixel photometers such as \textit{Kepler} and TESS and, in the future, PLATO. In particular, multi-color high-spatial resolution imaging was the first hint toward ruling out the planet scenario in this system. }

   \keywords{(Stars:) binaries: eclipsing, brown dwarfs, close -- Stars: oscillations}

\titlerunning{KOI-3886: the planet impostor}
\authorrunning{Lillo-Box et al.}

   \maketitle
%

\section{Introduction}

The complex process of detecting extrasolar planet candidates and validating their true planetary nature has been refined during recent years of exoplanet exploration. The launch of the \textit{Kepler} mission \citep{borucki10} and the many planet-like signals detected in its first months of operations made it necessary to come up with dedicated follow-up observing programs performing specific observations to rule out other possible configurations mimicking a planet-like transit signal. Specifically, high-spatial resolution images became a key starting point in this validation process \citep{howell11,law13,lillo-box12,lillo-box14b,Furlan17}. Given the large pixel size of the \textit{Kepler} CCD ($4\times4$ arcsec) and now even more importantly in the Transiting Exoplanet Survey Satellite (TESS; \citealt{ricker14}) photometer ($21\times21$ arcsec), understanding and measuring the flux contamination from both chance-aligned and bound sources are essential to validate planet candidates (especially the smallest planets, e.g., \citealt{barclay13}) and to refine their physical and orbital parameters \citep{lillo-box14b}. A second step in this process is the radial velocity (RV) monitoring of the host candidate to detect the amplitude and shape of the modulations to infer a mass for the transiting companion. Although this is not always possible given the low planetary masses (e.g., in the rocky regime, \citealt{barclay13}), high-resolution spectra can be used to provide upper limits to their masses and to discard additional scenarios undetectable by the high-spatial resolution images (e.g., close bound binaries) through analysis of the cross-correlation function (hereafter CCF). This proved to be especially relevant in some specific cases of planet-like signals that were subsequently uncovered to be of stellar or substellar nature (e.g., \citealt{santos02}). All this experience from exoplanet detection and follow-up efforts puts on the table the need for dedicated observations after transit measurements. 

In this work we describe a new example of an impostor planet. The \koi{} system showed all apparent signs of being a truly high-impact discovery with a potential gigantic planet around an evolved red giant branch (RGB) star skimming the Roche Lobe overflow point. {This gigantic planet would have had} an expected remaining lifetime of below five million years - a really lucky catch of the last breath of a monster planet. In this paper, we show evidence that leads to disguising KOI-3886.01 as a planet candidate in favor of a complex hierarchical triple system composed by a $\sim$K2~III star ascending the RGB orbited by an evolved $\sim$G4~IV subgiant star at a projected distance of 270 au, the latter surrounded by an eclipsing inflated brown dwarf in a very close-in orbital period of 5.6 days.

Brown dwarfs are substellar objects that do not have enough mass to trigger the burning of hydrogen in their nuclei. Theoretical models for their formation and evolutionary structure (i.e., mass, radius, luminosity) are difficult to calibrate given the very low number of brown dwarfs that have been fully characterized (i.e., have had their mass, radius, and age measured). To retrieve the size of these bodies and thus feed these evolutionary models, detecting eclipsing brown dwarfs is critical. So far, to our knowledge, only 26 eclipsing brown dwarfs have been detected (see \citealt{carmichael20} and references therein, as well as the latest discoveries by \citealt{jackman19,subjak20,palle21}). Most of these eclipsing   brown dwarfs are orbiting mature main-sequence stars of different spectral types, whose ages are difficult to determine using traditional isochrone fitting approach. Theoretical studies suggest that these substellar objects present large radii during the first stages of their formation while they are contracting to reach Jupiter-like (or even smaller) sizes. The very few eclipsing brown dwarfs detected around young stars (e.g., \citealt{stassun06}) show radii larger than the theoretical values, thereby pointing to this scenario. However, more mature eclipsing brown dwarfs also show radii larger than the expected by theoretical predictions. This can be attributed to stellar irradiation \citep{santisteban16} by the companion. However, the lack of precise age measurements also prevents a direct comparison with models. The \koi{} system offers an excellent environment in which the presence of a RGB star presenting solar-like oscillations allows a precise measurement of the age of the system. This measurement, together with the eclipsing nature of the brown dwarf, allows for a full characterization of its properties. Hence, \koi{} is a benchmark system for the study of brown dwarf properties in a hostile environment. 

Interestingly, the subgiant nature of the host star and the mass of the eclipsing object enhance the presence of the well-known reflexion, ellipsoidal, and beaming light curve modulations, which have recently been used to confirm (i.e., obtain the mass of) new planets (e.g., \citealt{lillo-box14,millholland17,lillo-box21}) and substellar objects (e.g., \citealt{mazeh12,lillo-box16a}) directly from a light curve as an alternative to RV monitoring.

The paper is organized as follows: in Sect.~\ref{sec:observations} we present the observations that lead to unveiling the nature of the KOI-3886 system; we then show evidence against the planetary nature of the eclipsing candidate in Sect.~\ref{sec:impostor}. In Sect.~\ref{sec:system} we use all the observations to fully characterize the system. We discuss the results in Sect.~\ref{sec:discussion} and conclude in Sect.~\ref{sec:conclusions}.

\section{Observations}
\label{sec:observations}

\subsection{Kepler and TESS photometric time series}
\label{sec:kepler}

\begin{figure}
\centering
\includegraphics[width=0.5\textwidth]{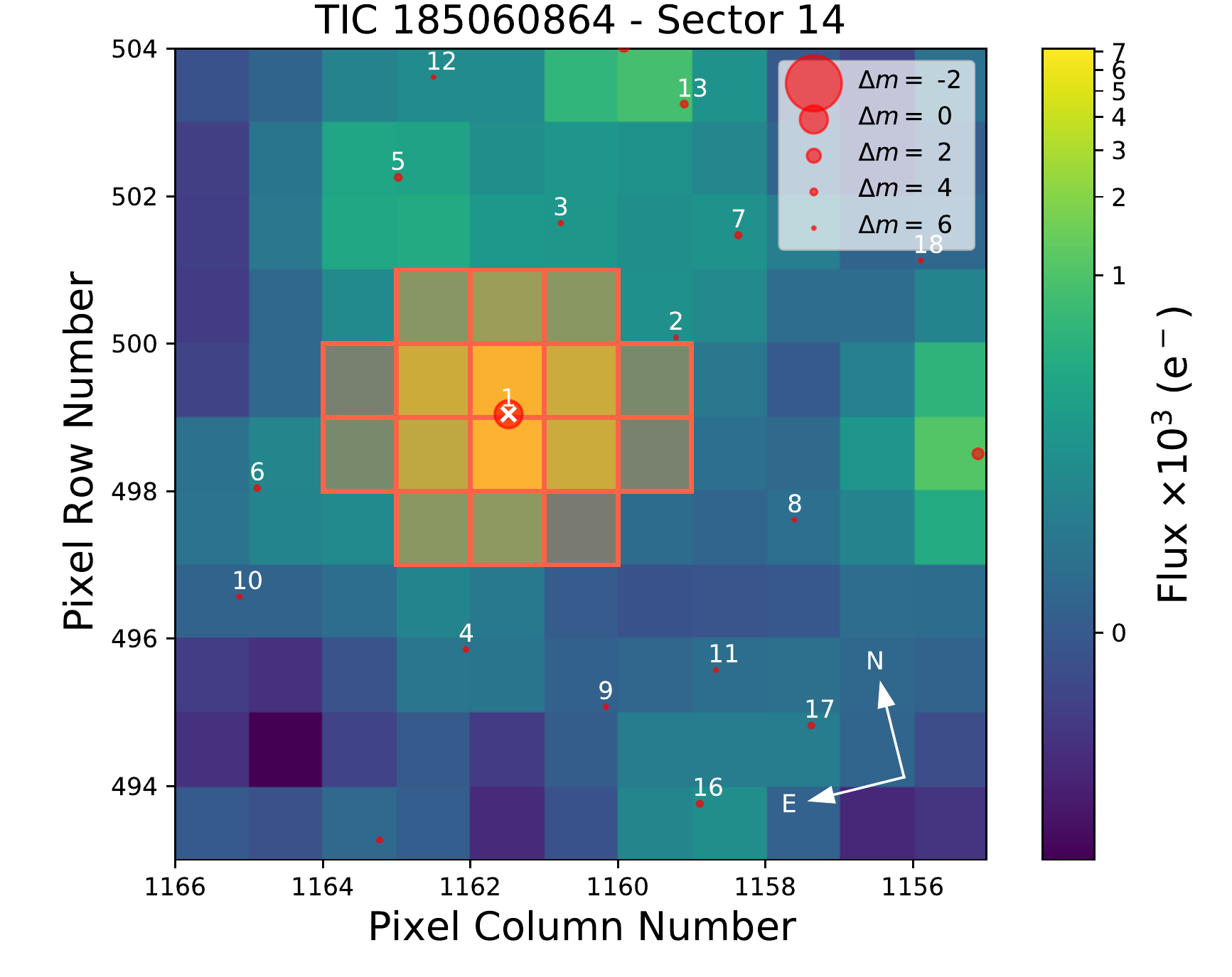}
\caption{Target pixel file of KOI-3886 obtained with \texttt{tpfplotter} \citep{aller20} from Sector 14 of the TESS mission.}
\label{fig:tpf}
\end{figure}

KOI-3886 (HD190655, KIC 8848288, TIC 185060864, see Table~\ref{tab:basic}) was observed by the \textit{Kepler} mission \citep{borucki10} between 2010 and 2014 in long-cadence mode, obtaining one photometric measurement every 30 minutes. The TESS mission also observed this target as part of its 2 min cadence legacy in Sector 14 during 27.4 days. Figure \ref{fig:tpf} shows the median target pixel file (TPF) from this TESS observation (computed using \texttt{tpfplotter}; \citealt{aller20}). The figure shows no contaminant sources within the aperture detected by Gaia DR2 down to 8 magnitudes contrast against the Gaia magnitude of the target. The same applies to the \textit{Kepler} aperture. 

\begin{table}[]
\setlength{\extrarowheight}{3pt}
\caption{\label{tab:basic} General properties of the KOI-3886 system.}
\begin{tabular}{lll}
 
 \hline 
Parameter & Value & Ref.$^{\dagger}$ \\ 
\hline 
IDs & KOI-3886, HD\,190655, \\
        &  KIC\,8848288, TIC\,185060864 \\
Gaia EDR3 ID & 2082133182277361152  & [1] \\
RA, DEC & 20:04:11.33, +45:05:15.30  &  [1]\\ 
Parallax (mas) & $2.14 \pm 0.30 $    & [1]\\
$\mu_{\alpha}$ (mas/yr) & $-8.41\pm 0.37$  & [1]\\
$\mu_{\delta}$ (mas/yr) & $-3.18\pm 0.37$  & [1]\\
RV (km/s) & $-24.00 \pm 0.64$  &  [1] \\ 
G (mag) & 10.1  &  [1]\\
$B_p-R_p$ (mag) & 1.28  &  [1]\\
m$_V$ (mag) & $10.306 \pm 0.050$ &  [2]\\
Ks (mag) & $7.475\pm0.023$ & [3] \\
U (km/s) & -3.955 & [4]\\
V (km/s) & -23.478 & [4] \\
W (km/s) & -3.026 & [4]\\
Gal. population & Thin disk & [4] \\

 \hline

\end{tabular}
\tablebib{
[1] Gaia Collaboration et al. (2016); 
[2] Bessell et al. (1998); 
[3] Cohen et al. (2003); 
[4] This work.}
\end{table}

\begin{figure}
\centering
\includegraphics[width=0.5\textwidth]{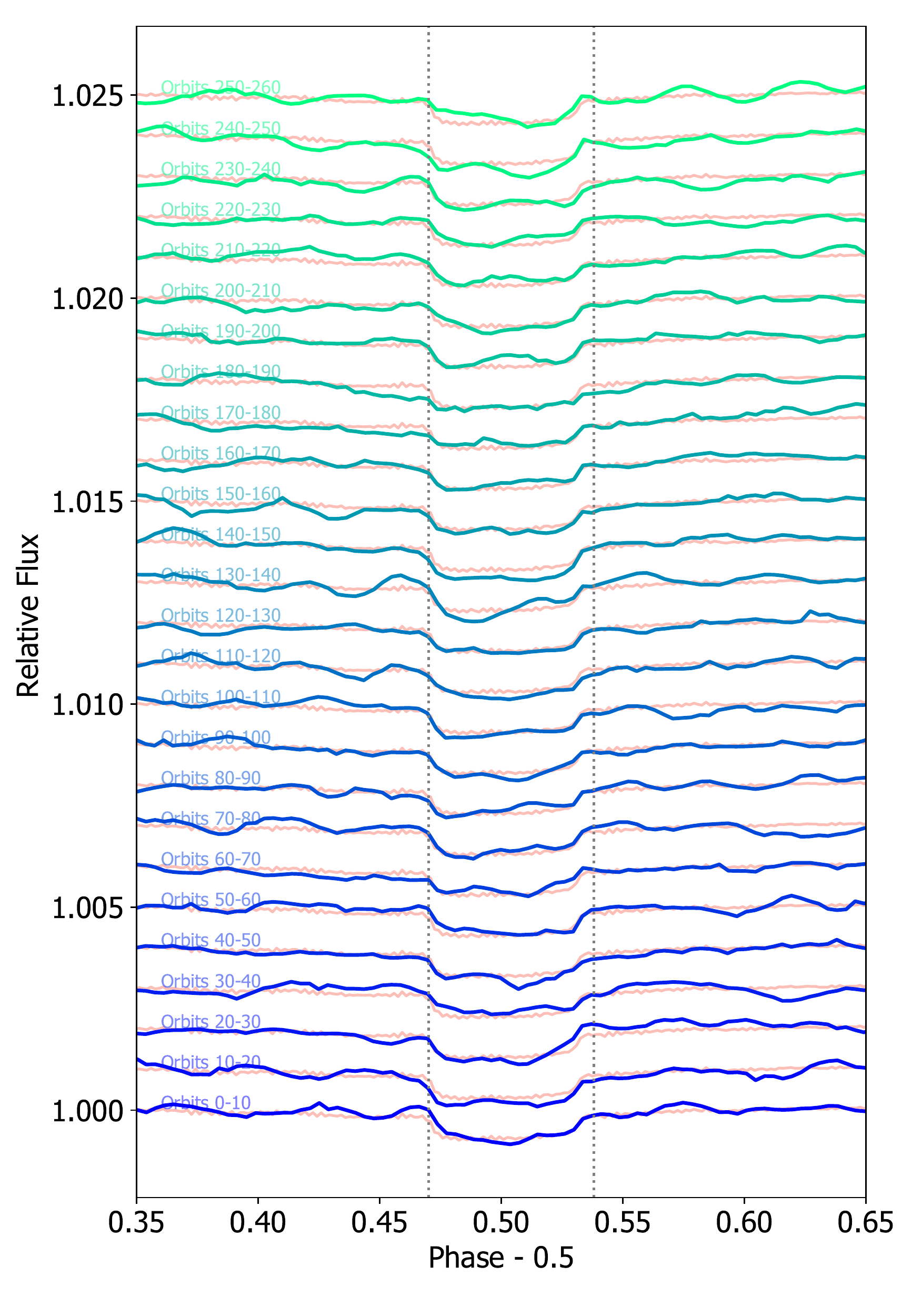}
\caption{Phase-folded \textit{Kepler} light curve centered in the mid-eclipse. Each line includes a combination of 10 orbits. The asteroseismology signal from \kA{} is mixed with the eclipse signal at a similar flux variation level. The full combined light curve is shown for reference on each chunk.}
\label{fig:secular}
\end{figure}

\begin{figure*}
\centering
\includegraphics[width=1\textwidth]{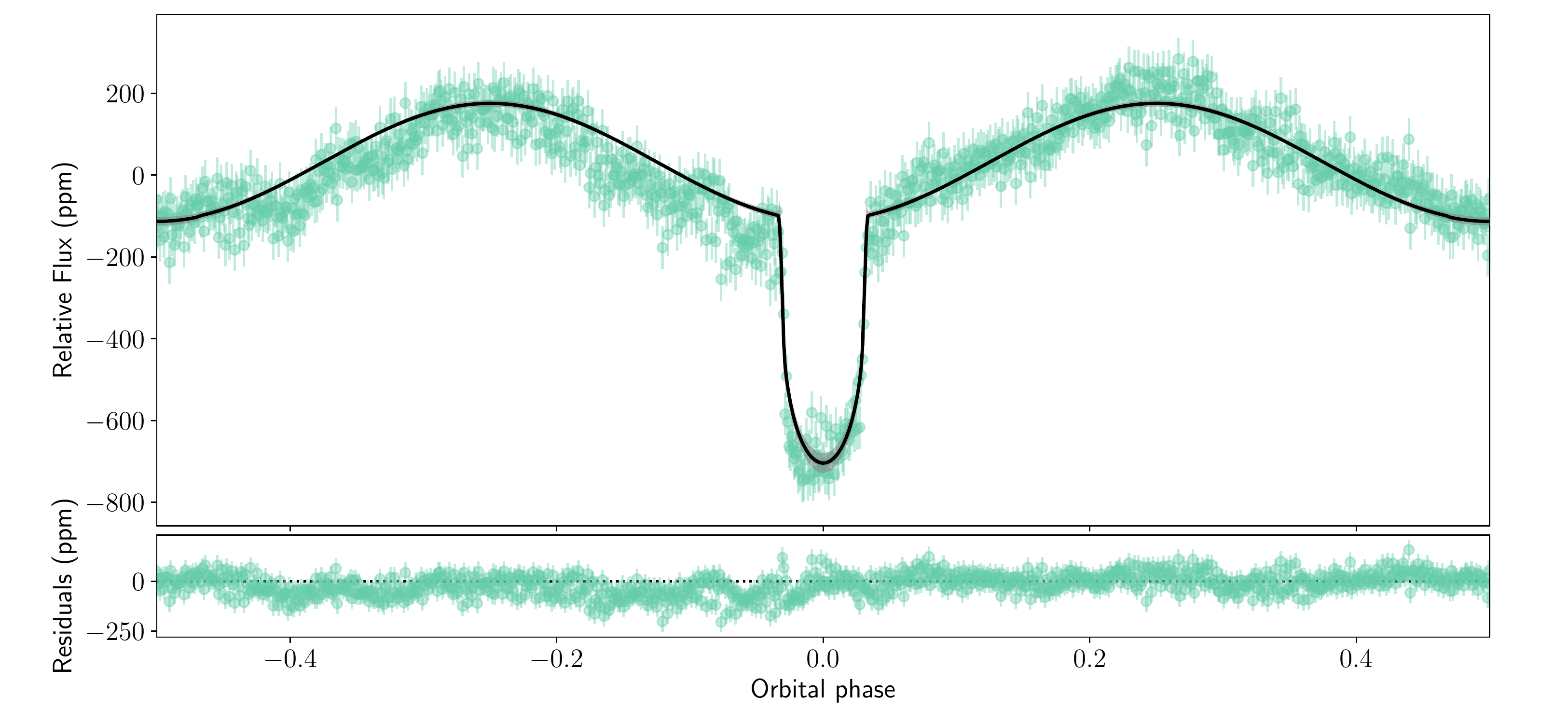}
\caption{Diluted \textit{Kepler} light curve phase folded with the eclipse period and the inferred median model from the joint analysis from Sect.~\ref{sec:joint}. The corresponding 95\% confidence interval is shown as a vgray shaded region. \textit{Kepler} data has been binned in phase to remove the asteroseismology variation due to the contaminant bright \kA{} star.}
\label{fig:LCphase}
\end{figure*}

We obtained the \textit{Kepler} and TESS light curves from the MAST archive\footnote{\url{https://mast.stsci.edu/portal/Mashup/Clients/Mast/Portal.html}} and used the PDCSAP (Pre-Data Conditioned Search Aperture Photometry) flux for our analysis. The periodogram of the \textit{Kepler} light curve shows evidence for a modulation at 5.56 days, corresponding to the transiting candidate planet KOI-3886.01 reported by the \textit{Kepler} team \citep{batalha13}. In Fig.~\ref{fig:secular}, we show the combination of chunks of 10 orbits per line in the region around the detected eclipse signal. In total, more than 260 orbits are available in the \textit{Kepler} data. When phase-folding the light curve with such periodicity we noticed two clear signals: the transit dimming and a clear double hump in the out-of-transit region (see Fig.~\ref{fig:LCphase}). The transit signal has a depth of about 700 ppm (parts per million). The out-of-eclipse modulations indicate the presence of ellipsoidal variations due to tidal interaction between the star and the transiting object. The amplitude of these modulations is at the level of 200 ppm.  

\begin{figure}
\centering
\includegraphics[width=0.5\textwidth]{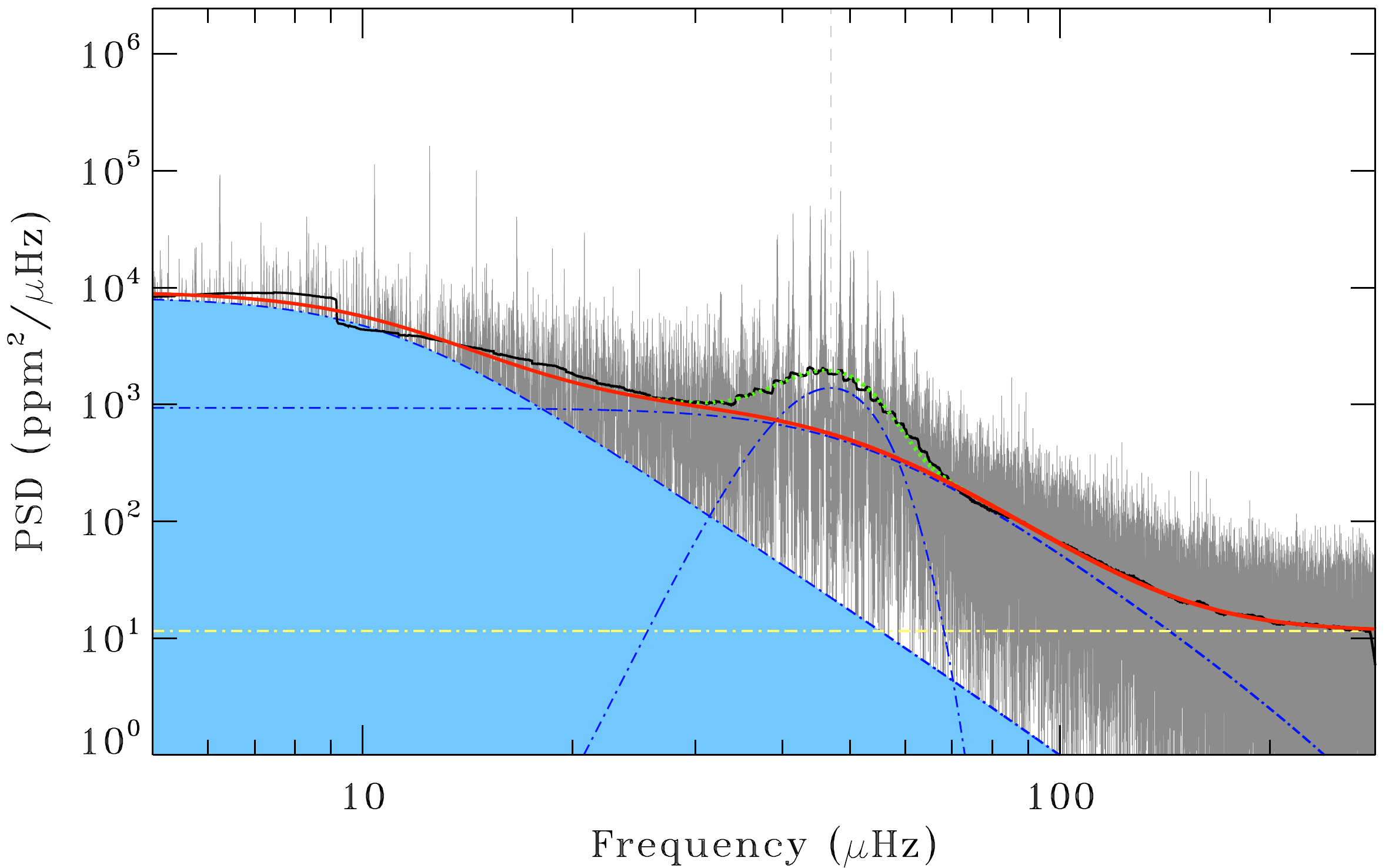}
\caption{Power spectrum of KOI-3886. The power spectrum is shown in gray (with a heavily smoothed version in black). The solid red curve indicates a fit to the background done with \texttt{DIAMONDS} \citep{corsaro14}, consisting of two Harvey-like profiles (dot-dashed blue curves) plus white noise (horizontal dot-dashed yellow line). A joint fit to the oscillation power excess (dot-dashed blue Gaussian curve) and background is visible at $\sim50\:\mu\rm{Hz}$ as a dotted green curve.}
\label{fig:background}
\end{figure}

The periodogram also shows a clear forest of power frequencies around 50 $\mu$Hz (see Fig.~\ref{fig:background}), corresponding to typical solar-like oscillations of a red giant star (as shown for instance in the case of the confirmed planet host Kepler-91; see \citealt{lillo-box14}). This is analysed in depth in Sect.~\ref{sec:asteroseismology}. 

The \textit{Kepler} Data Validation Report (DVR) indicates large excursions of the out of transit centroids ($>$  3 arcsec from the target position). However, these are probably caused by the large amplitudes of the asteroseismology oscillations, implying flux variations of up to 1 part per thousand (ppt). 

Given the largely shorter temporal coverage of the TESS data set against the \textit{Kepler} data (27.8 days versus 1400 days) and the presence of the above described stellar oscillations, which make it necessary to combine a large set of orbits to average out this effect, we decided to not use the TESS dataset further for the sake of simplicity in this paper.

\subsection{High-spatial resolution imaging}
\label{sec:imaging}

\begin{figure}
\centering
\includegraphics[width=0.5\textwidth]{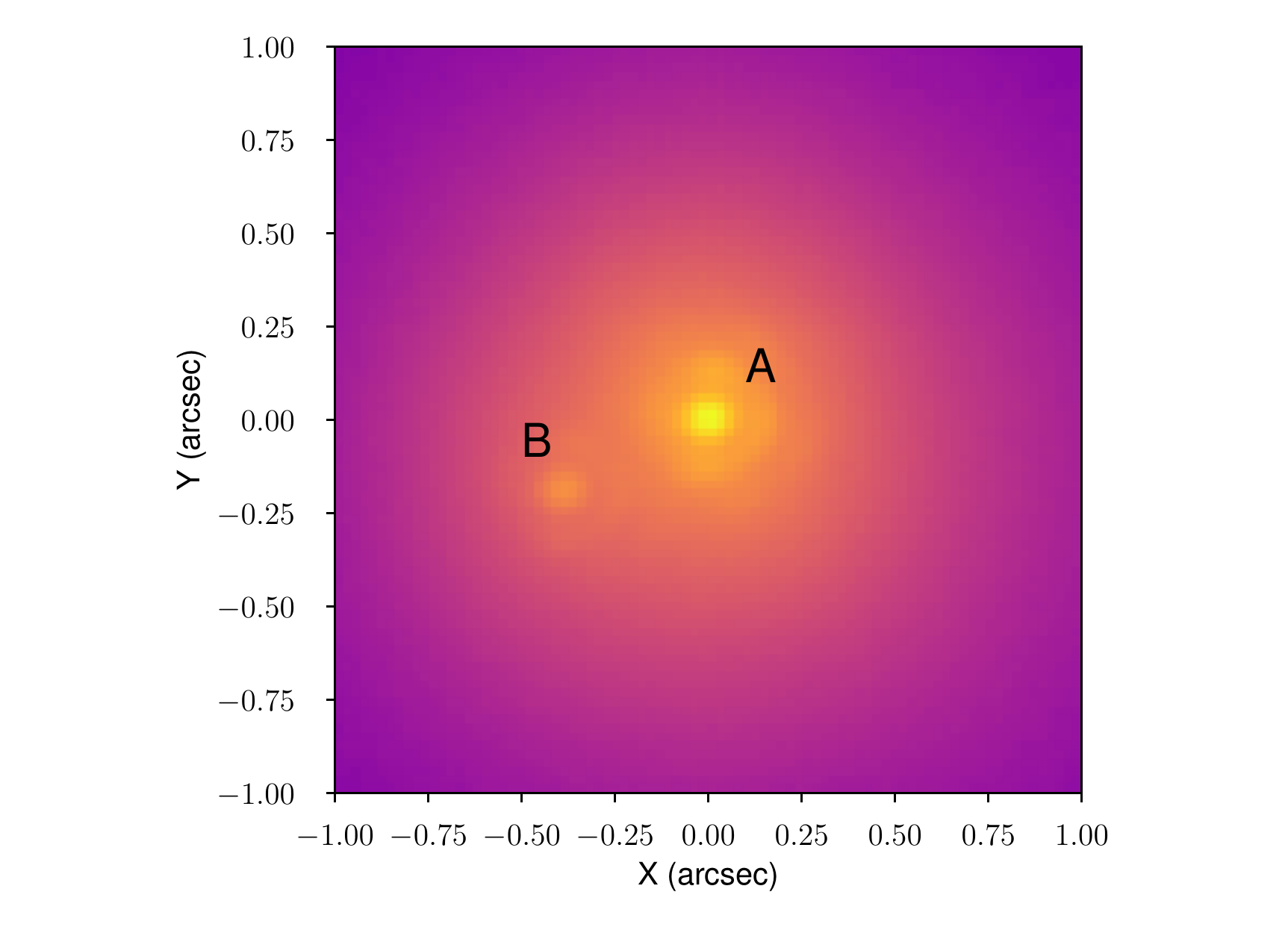}
\caption{AstraLux image of the KOI-3886 system.}
\label{fig:astralux}
\end{figure}

In \cite{lillo-box14b}, we presented a high-spatial resolution image of \koi{} obtained through the lucky imaging technique with the AstraLux North instrument \citep{hormuth08} located at the 2.2 m telescope at the Calar Alto Observatory (Almería, Spain). The image (see Fig.~\ref{fig:astralux}) showed a very close companion (hereafter called \kB{}) located at 0.43 arcsec to the brightest object (\kA{}) and with a magnitude contrast of 0.85 mag in the $i^{\prime}$ band (0.99 mag in the $z^{\prime}$ band). Additionally, images from Robo-AO \citep{ziegler17} and PHARO \citep{Furlan17} were also published in the literature. Both high-resolution images also detected this close companion with a magnitude contrast of 2.2 mag in the K band. Based on these contrast magnitudes we can estimate a dilution factor of 43\% in the \textit{Kepler} band, assuming the eclipses occur in the brighter companion. This is a significant source of contamination, which needs to be taken into account in the modeling of the light curve. It also brings up the question of which is the host of the eclipsing object. If star B is the host, the dilution factor becomes 71\%.

\subsection{Spectral energy distribution}
\label{sec:SEDobs}

Table \ref{Table:PHOT} shows the observed photometry of KOI-3886. Among the large amount of observations available, we carried out a careful and critical selection of the magnitudes, therefore the values listed in the table can be considered as a clean set of data. Sources of fluxes at zero magnitude are as follows: Tycho $BV$ \citep{Mann15}, Johnson $UBV$ \citep{Bessell98}, and 2MASS $JHK_{\rm s}$ \citep{Cohen03}. The GALEX NUV \citep{Olmedo15} and Sloan magnitudes $g'r'i'z'$ \citep{Fukugita96,York00} are on the AB system, thus a flux of 3631 Jy was adopted in both  cases. Regarding the GALEX NUV flux, which plays an important role  in the spectral energy distribution (SED), discrepancies between the filter effective wavelength are found in the literature. The value of the GALEX NUV\ flux affects the  conversion of the fluxes in Jy, computed from the AB magnitude, to cgs units used in this work;   in this work we adopted the value 2315.7~\AA{} from \citet{Beitia16}.
  
\begin{figure}[!tbh]
\centering
\includegraphics[width=0.5\textwidth]{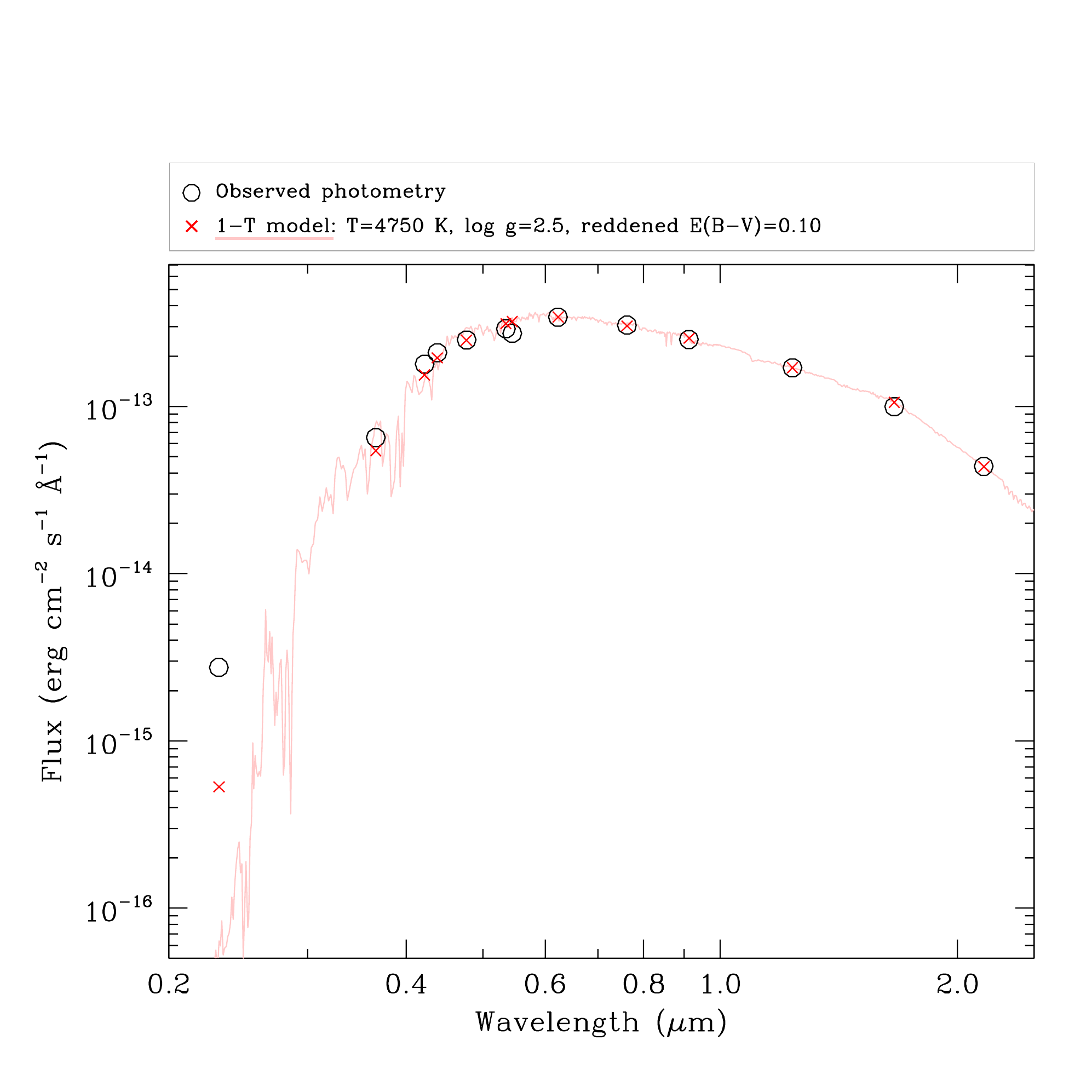}
\caption{Observed photometry of KOI-3886 (empty black circles), and a 1T-model fit of a Kurucz model \citep{kurucz79} with $T_{\rm eff}$=4750 K and $\log g_*$=2.5 (red crosses and solid pale red line). Whereas  the red part of the SED is well reproduced by the model, the fluxes below $\sim\!5000$ \AA, and in particular that corresponding to the NUV GALEX band, are underestimated.}
\label{Fig:SED-ONE-T}
\end{figure}

The observed SED is shown in Fig. \ref{Fig:SED-ONE-T} as empty black circles. A one-component temperature fit, plotted as red crosses and a pale red line, shows that whereas the photometry at optical ($\lambda\!>\!5000$~\AA) and near-infrared wavelengths is well reproduced, an additional hotter component, not modifying the shape of the final, composite fit at these wavelengths, but accounting for the deficit of fluxes at bluer wavelengths, and especially for the NUV GALEX point, is required. This clearly suggests that the brightest component (\kA{}), which dominates the region above $\sim$5000 \AA, must be colder than the fainter component (\kB{}). This is very relevant to unveil the architecture of this system. The composite fit to this SED is shown in Sect.~\ref{sec:SED}.


\subsection{High-resolution spectroscopy and preliminary radial velocities}
\label{sec:HRspectroscopy}

We observed \koi{} using the fiber-fed high-resolution spectrograph HERMES \citep{raskin11} located at the 1.2 m Mercator telescope at the Roque de los Muchachos Observatory (La Palma, Spain). The observations were carried out during five consecutive nights on 15-19 July 2020 in visitor mode and another five epochs within a week time span between 22-29 July 2020. We used the simultaneous reference observing mode, in which the second fiber is illuminated by ThAr and Ne lamps, which imprints the emission lines from these species in the inter-order region of fiber B. This strategy allows us to keep track of intra-night RV drifts at the 2 m/s level. Added to this, we observed the RV standard HD 109358 (V=4.25 mag) to measure possible night-to-night offsets since the instrument is not pressure-stabilized (only the temperature is actively controlled). Among the five nights, the third was performed under poor weather conditions with high clouds interrupting the observations. Hence, the data from that night are of less quality. We obtained three spectra per night on the first run and one spectrum per night on the second run. The typical signal-to-noise ratio (S/N) of the dataset is 20. The data were reduced by using the standard pipeline of the instrument. We first extracted the RV through cross-correlating the spectra with a binary mask constructed from a solar-like spectrum. We measured the nightly offsets by using the RVs of the standard star. This nightly offsets were then subtracted from the target RVs. The median precision on the RVs is 5.6 m/s.

High-resolution spectra of \koi{} were also obtained with the CAFE fiber-fed spectrograph \citep{aceituno13,lillo-box19} installed at the coud\'e room of the 2.2 m telescope of the Calar Alto Observatory (Almería, Spain). The system was observed on the nights of 23-24 of July 2020, together with the RV standard HD\,109358 to measure inter-night RV offsets since the instrument is not pressure-controlled. Additionally, we obtained ThAr reference frames before and after each science image to measure the intra-night drift. The conditions on both nights were optimal with a typical seeing of 1 arcsec. We obtained 18 spectra of \koi{} (three per night spread along the whole night) with a mean S/N of 30. We reduced the spectra using the instrument pipeline \texttt{cafextractor} \citep{lillo-box19}. Besides the basic reduction and the wavelength calibration, \texttt{cafextractor} measures the RV of each spectrum by cross-correlating them with a G2 mask custom-made for the instrument. The pipeline uses the drift measurement to correct this RV. Additionally, we remove the offset corresponding to the RV standard as measured at the beginning of the night. The final median photon noise of the RVs corresponds to $\sim$10 m/s. 

\subsection{Gaia}
\label{sec:gaia}

We checked the \koi{} system on the Gaia EDR3 catalog \citep{lindegren20}. Although this target was not reported in the second Gaia release (DR2), the EDR3 release now includes this target (Gaia EDR3 2082133182277361152). The catalog provides a parallax of $\pi=2.13\pm0.31$~mas and proper motions of $\mu_{\alpha}=-8.41\pm 0.37$~mas/yr, $\mu_{\delta}=-3.17\pm0.37$~mas/yr (see Table~\ref{tab:basic}). {The corresponding UVW galactic velocities indicate this source probably belongs to the thin disk \citep{bensby03}}. The catalog does not include star B, which hence acts as a contamination source to the reported information on this system. The Gaia data provides a large excess noise ($\epsilon i = 3.1$) and a significance of the excess noise of $D>10000$, which is an extremely large value. Also, the renormalized unit weight error (RUWE) parameter of $\sim30$ is significantly larger than the threshold value of 1 expected for sources for which the single-star model provides a good fit to the astrometric observations \citep{lindegren20}. This is compatible with the presence of \kB{} and may be influenced by the asteroseismic modulations in \kA{} and the eclipse variations detected in the \textit{Kepler} light curve. Additionally, the Gaia catalog indicates the presence of another two close-by sources (Gaia EDR3 2082127547280244224 and Gaia EDR3 2082127478555928448) located $\sim$1.7~arcmin from the \koi{} system. According to the EDR3 catalog, these sources are both at around 650~pc (one of which is possibly an M dwarf), with very similar proper motions as \koi{} ($\mu_{\alpha}=-10.2$~mas/yr, $\mu_{\delta}=-2.8$~mas/yr and $\mu_{\alpha}=-6.5$~mas/yr, $\mu_{\delta}=+3.4$~mas/yr). This distance, as shown in Sect.~\ref{sec:discussion}, is perfectly compatible with that estimated for our target system. The projected separation corresponds to 74\,000 au, hence one-fourth of the Sun-Proxima Centauri distance. However, since these targets are not included in the \textit{Kepler} aperture we do not consider them in the subsequent analysis. But we want to highlight that their presence could make this system even more interesting dynamically speaking.

\section{An impostor planet}
\label{sec:impostor}

\subsection{The planet-like signal}
We used the preliminary RVs derived in Sect.~\ref{sec:HRspectroscopy} to perform a first stand-alone analysis of the data. The RVs show a clear modulation with the same periodicity as the transit signal seen in the \textit{Kepler} and TESS data. We first modeled the dataset using a one-Keplerian model with a uniform prior on the orbital configuration (eccentricity and argument of the periastron) and RV semi-amplitude. The tidal modulation induced by the planet candidate into the stellar photosphere also creates an ellipsoidal shape in the star that affects the RV. We accounted for this effect by adding a second term to the RV signal as in \cite{arras12}, which has a periodicity corresponding to twice the orbital period. We used a uniform prior between 0 and 100 m/s for the semi-amplitude of the tide. The parameter space was explored using a Markov Chain Monte Carlo (MCMC) scheme using the \texttt{emcee} implementation (\citealt{emcee}). We used four times the number of parameters as walkers ($N_w =40$) and 20\,000 steps per walker. After a first exploration of the parameter space, we performed a burn-in phase and then focused on a small region around the maximum-likelihood solution to obtain reliable uncertainties to our parameters. The solution does not constrain the eccentricity of the orbit of the planet. Hence, we proceeded with a pure circular orbit for the planet. The solution provides a larger Bayesian evidence (estimated through the \texttt{perrakis} code, \citealt{perrakis14}) for the circular model. The results of this model show a Keplerian signal with a semi-amplitude RV of $K=150.8\pm7.2$ m/s. The tidal effect induced a clear signal of  $K_{\rm tide}=15.2 \pm 7.3$~m/s. Assuming the stellar properties published in the \textit{Kepler} Input Catalog \citep{brown11}, the Keplerian signal corresponds to a minimum mass for the planet candidate of $m_p\,\sin{i} = 1.83\pm0.14$~\Mjup{}.

Separately, we also explored the \textit{Kepler} light curve in a stand-alone fashion. The light curve from KOI-3886 contains several effects that deserve preliminary qualitative discussion. The out-of-eclipse light curve shows a double hump typical of ellipsoidal modulations. Both peaks are located at quadrature phases ($\phi=0.25$ and $\phi=0.75$), as expected for a circular or near-circular orbit. Also, their amplitudes are slightly different, hence we expect the Doppler beaming to play a role. The flux at mid-phase ($\phi=0.5$) is compatible with the flux just before or after the transit, hence the reflection component seems negligible. Additionally, stellar pulsations are evident in the light curve and even clearer in the periodogram. Their amplitudes are relatively large compared to the ellipsoidal amplitude and transit depth. Hence, the detection of individual transits is blurred by these oscillations. The transit signal in the phase-folded light curve (where pulsation effects are averaged out) shows a clear asymmetric shape, with the egress longer than the ingress. This asymmetry can be interpreted as a non-zero orbital eccentricity (e.g., \citealt{kane08}). However, the difference in duration of both effects would imply an eccentricity that is too large and incompatible with the RV data and the dynamical analysis of the system, which should have circularized the orbit already (unless it is currently decaying fast to the host star). Alternatively, the asymmetry could come from an extended cometary-like tail as in the case of evaporating planets (e.g., \citealt{bourrier14}). A third possibility would be the tidal effect exerted by the star, which would make the planet shape to stretch in the planet-star direction. However, in a circular orbit, this would lead to symmetric ingress and egress shapes. A small but not null eccentricity might explain the asymmetry.

In order to average out the large photometric variations due to the stellar pulsations, we phase-folded the \textit{Kepler} light curve with the eclipse period and binned the dataset in phase to reach 1\,000 data points. This means an average binning of 67 data points per bin in phase. With this approach we effectively removed the large asteroseismic variations. We then proceeded with a simple analysis of the whole \textit{Kepler} dataset including the reflexion, ellipsoidal, and beaming (hereafter REB) modulations and the transit and eclipse model. For the REB modulations, we used the equations and parametrization from \cite{lillo-box14} and for the transit and eclipse signals we used the \texttt{batman} code \citep{kreidberg15}. We used \texttt{emcee} to model the data, and used a prior for the inclination that ensured a non-grazing transit. The posterior distribution of the parameters converged appropriately except for the inclination, which tends to get below the allowed range toward more inclined values implying a grazing eclipse. The results for the planet mass and radius from the light curve modeling show a strong degeneracy between different parameters, but pointed to a large $R_p>2.5~R_{\rm Jup}$ and massive ($>1~M_{\rm Jup}$) planet, in relative agreement with the RV analysis.

Based on these results, this planet would have been the closest planet to a red giant star and its semimajor axis placed it as the closest planet to the Roche Lobe with $a/a_{\rm RL}=1.2$, twice closer than the closest ones (Kepler-91\,b at $a/a_{\rm RL}=2.3$; \citealt{lillo-box14}- and K2-141\,b $a/a_{\rm RL}=2.2$; \citealt{malavolta18}). In terms of planet evolution, this would have been a major discovery, as we would be watching the planet death in real time. In this scenario (eclipsing object around \kA{}), we can compute the dynamical evolution of the system as the star leaves the main sequence following \cite{Villaver09} and \cite{Villaver14}. We find  that using a standard stellar evolution model and the measured parameters (an almost circular orbit with an initial period of 5.56 days) the planet would not have survived to the point at which the star reaches the current stellar radius of 11.2 R$_{\odot}$. Instead, it would have been engulfed before, at approximately R$_{\star,A}$ = 8 R$_{\odot}$, as a result of tidal forces. The planet would have fallen into the stellar envelope in only $\sim$6~Myr. Such a short scale of evolution makes it very unlikely to observe phenomena like this. To force its survival to the current stellar radius-orbit ratio we have to use a larger initial orbit (0.1 au versus the 0.07 au) or force the standard tidal parameters to such values that basically inhibit tidal decay as the star evolves off the main sequence. All this made the planet scenario around \kA{} very challenging and unlikely, pointing to alternative solutions.

\subsection{The false positive detection}
\label{sec:FP}

Despite the clear signs that the eclipsing object has a planetary nature in both the RV and photometric time series, the presence of a very close companion to the main target and its relatively small contrast made it necessary to analyze the data more in depth. The  CCF for all spectra only shows one apparent sharp component, suggesting that only one of the two sources inside the fiber possesses a spectrum compatible with the lines in the binary mask, that is, a solar-like spectrum. However, we also estimated the bisector of the CCF to check for potentially blended components. By using the CCFs calculated with the G2 mask, we found a strong correlation of the bisector with the RVs, indicating that some deformation of the CCF profile was the responsible for the RV variations (see Fig.~\ref{fig:colorRVs}). The correlation with the bisector is an indication of potential blending scenario similar to that observed in the HD\,40114 system \citep{santos02}. Subsequently, we recomputed the CCF using different binary masks, namely from spectral types F9, G2, K2, and M0. We then computed the RVs from the entire wavelength range and also by splitting it into three different color chunks (blue: 3800-5100~\AA, green: 5100-6400~\AA, and red: 6400-7800~\AA). In Fig.~\ref{fig:colorRVs} we show the results of this exercise. We can clearly see a dependence of the amplitude of the RV curve with the mask spectral type, increasing progressively toward bluer masks. Additionally, for a given mask, the bluer wavelength range produces larger amplitudes. All this is a clear indication of an additional hotter but shallower component in the CCF.


Consequently, we performed a two-component analysis of the CCF, including a sharp component and a broad shallower component (see an example in Fig.~\ref{fig:two-ccf}). We fitted all CCFs obtained with the G-type mask from the HERMES instrument simultaneously with the full width at half maximum (FWHM) of the Gaussian profiles ($\sigma_S$ - sharp - and $\sigma_B$ - broad) as common parameters for all epochs, but letting the contrast, center, and level of the profiles free for each different epoch. With 20 epochs, we used an MCMC approach to obtain the posterior distribution of all 123 parameters, including a jitter for the CCF for each epoch. 

\begin{figure}
\centering
\includegraphics[width=0.48\textwidth]{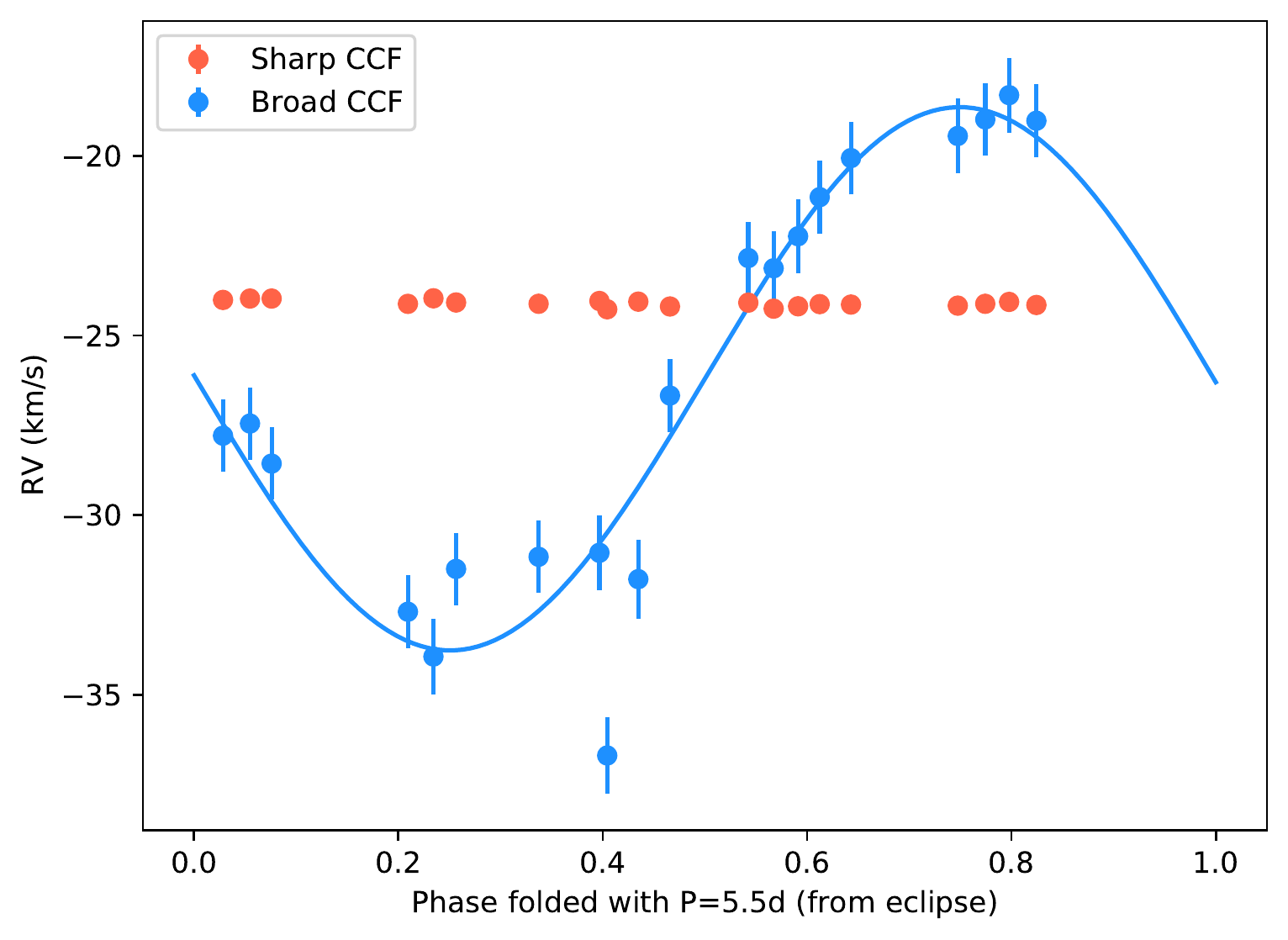}
\caption{Radial velocity variations of the two components of the CCF from the HERMES dataset phase-folded with the period from the eclipses seen in the \textit{Kepler} time series.}
\label{fig:2comp}
\end{figure}

Interestingly, the MCMC was able to recover the second, very shallow component (ten times shallower than the sharper component and three times broader). The center of the Gaussian profile for each of the components determines the RV after the barycentric Earth RV correction is applied. Figure~\ref{fig:2comp} shows both time series and the phase-folded RV curve assuming the period from the eclipse signal. From this figure, it becomes evident that the 5.6 day periodicity is actually occurring on the broad and shallow component, with a much larger RV semi-amplitude of $\sim$7 km/s. This blended component combined with the sharp component was actually producing a planet-like signal that resulted in a semi-amplitude of 150 m/s, which is compatible with a planetary-like signal.

We thus conclude that the RV variations and the eclipses seen in the \textit{Kepler} light curve are not due to a close-in planet but to an eclipsing binary whose CCF and light curve are diluted by a companion star. In the following sections we analyse the system taking this information into account to unveil which of the two stars (A or B) is the host of the eclipsing object and to characterize the properties of the system.

\section{System parameters}
\label{sec:system}

\subsection{Who is who?}
\label{sec:who}

\begin{figure*}
\centering
\includegraphics[width=1\textwidth]{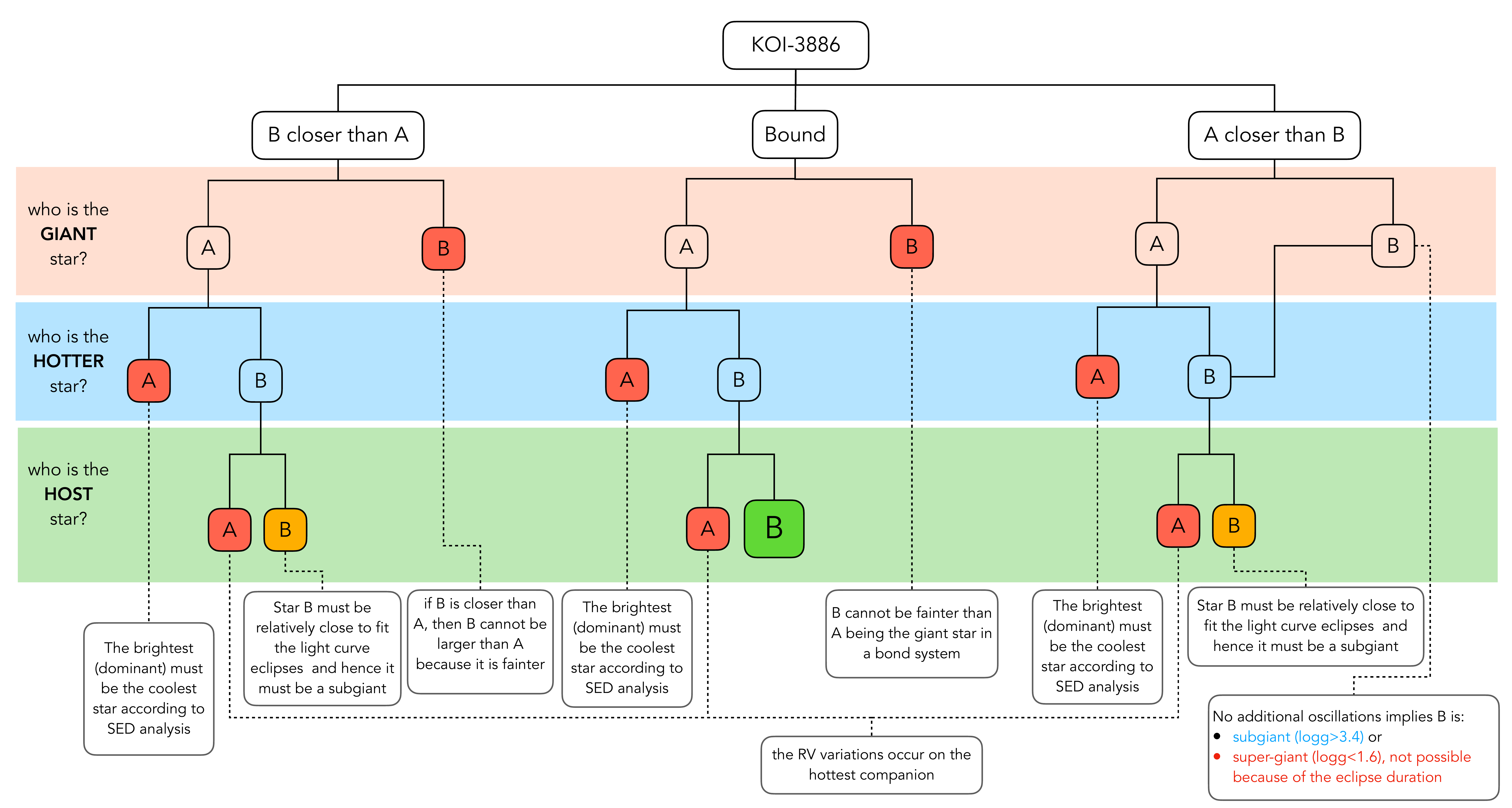}
\caption{Decision tree diagram to unveil the \koi{} system. The red boxes represent the invalid branches and the reasons are provided in the linked boxes using dashed lines. The green box is the preferred scenario owing to the probabilistic reasons provided in Sect.~\ref{sec:who}. The branches finishing in orange boxes cannot be discarded although they seem unlikely owing to these probabilistic arguments.}
\label{fig:who}
\end{figure*}

From the above analysis we have several clues to solve the \textit{who-is-who} in this system. We know that: 
i) there is an eclipsing system (Sect.~\ref{sec:kepler}), 
ii) there is only one giant star with $1.6<\log{g}<3.4$ since there is only one set of oscillation frequencies within the accessible range (see Fig.~\ref{fig:background} and Sect.~\ref{sec:kepler}), 
iii) there is a hot component and a cold component from the SED (see Sect.~\ref{sec:SEDobs} and \ref{sec:SED}), 
iv) the cold component is brighter and dominates the SED (Sect.~\ref{sec:SEDobs} and \ref{sec:SED}),  v) the hot component is the one hosting the eclipses (see Sect.~\ref{sec:FP}), and
vi) the star hosting the eclipses must be an evolved star ($R>2.3$~\Rsun{} for $M>1$~\Msun{}) to account for the amplitude of the ellipsoidal modulations displayed at the given periodicity. 
 
With all this information we build a decision tree to unveil the relative location of stars A and B to clarify the system architecture. This is shown in Fig.~\ref{fig:who} where we use the above arguments to discard the different possible scenarios. In summary, the only branches that we cannot discard based on the data indicate that star A is the giant star producing the pulsations and that star B is the star hosting the eclipses and should hence also be an evolved star. From these data, however, it is not possible to discern whether components A and B are actually bound or if they are simply a chance alignment, in   which case star B is slightly closer in the first branch, although it cannot be too far from star A because otherwise it would be a main-sequence star, which is incompatible with the ellipsoidal modulations at that periodicity; or star B could be farther away in the third branch. Equivalently, it cannot be {too} far away because otherwise it would be a super-giant star, which is incompatible with the ellipsoidal and RV data.

However, the unbound (chance-aligned) scenario in both cases seems unlikely because of stellar population statistics. In order to quantify this argument, we can estimate the probability of having two evolved stars at a separation below 1 arcsec in this region of the sky. We followed a similar procedure as that described in \cite{lillo-box14b} and also used in other works (e.g., \citealt{Themessl18}). Such probability is given by $P_{\rm chance}=\pi r^2 \rho_{\rm evolved}$, where $r$ is the maximum radius to compute the probability and $\rho_{\rm evolved}$ is the density of evolved stars in the region. Since the separation between star A and B is 0.43 arcsec, we can conservatively use $r=0.5$~arcsec. To estimate $\rho_{\rm evolved}$ we used the TRILEGAL\footnote{\url{http://stev.oapd.inaf.it/cgi-bin/trilegal}} galactic model \citep[v1.6,][]{girardi12} to retrieve a simulated stellar population around the location of the \koi{} system. We used the \texttt{astrobase} implementation \citep{astrobase} to simulate such population and compute the density of stars around the target position. We simulated stars up to magnitude $V=20$~mag and selected only those with $\log{g}<4$ to conservatively include evolved stars from the subgiant phase. By using this, we obtain a density of 6200 evolved stars per deg$^2$. By applying the above equation, we obtain a probability of chance alignment of two evolved stars within 0.5 arcsec to be 0.038\%. This can be interpreted as a probability of 99.96\% of the two evolved stars being actually bound.

\begin{figure}
\centering
\includegraphics[width=0.48\textwidth]{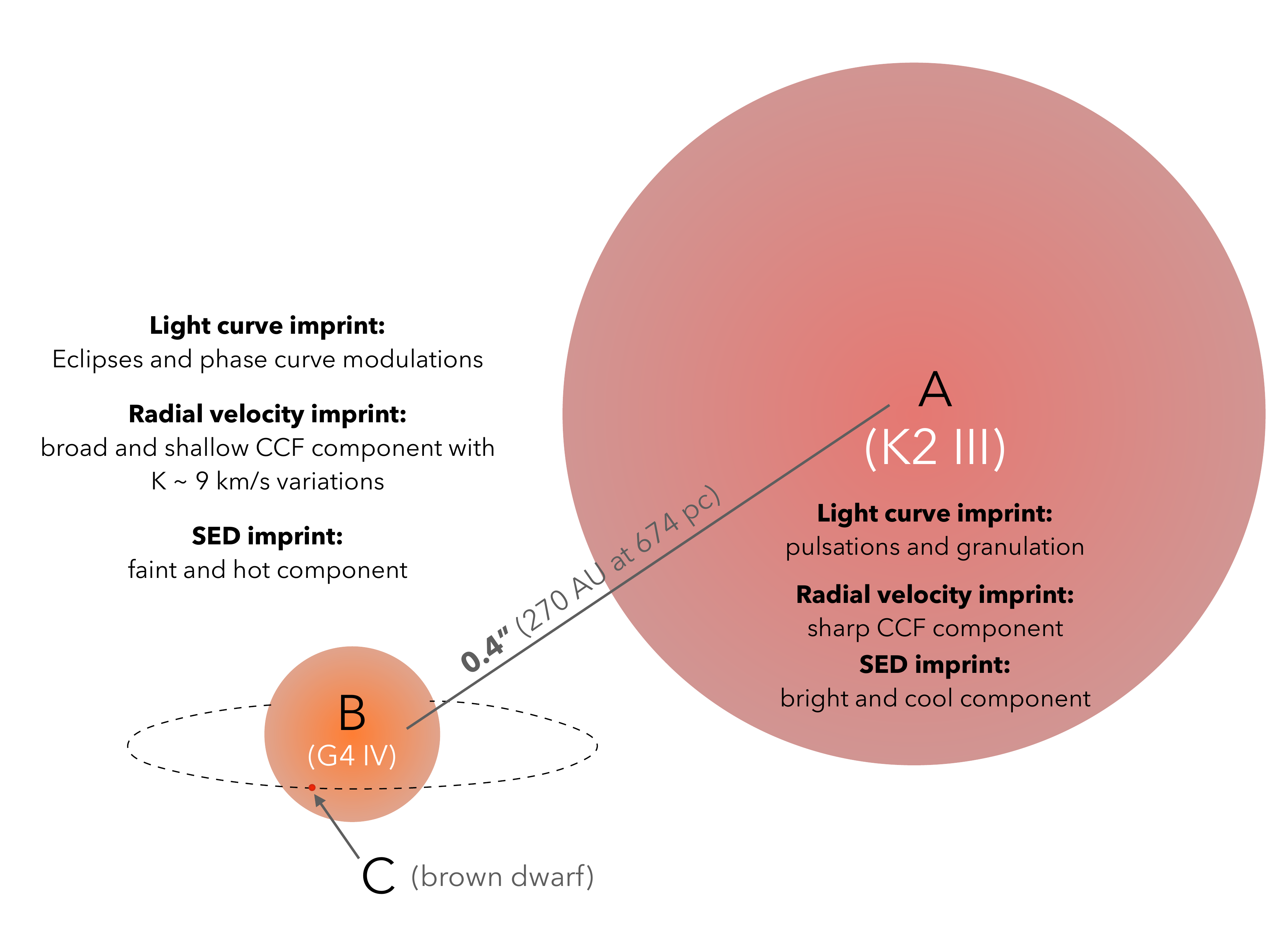}
\caption{Schematic view of the hierarchical triple scenario inferred for the KOI-3886 system; component A is a $\sim$K2 III giant star accompanied by an eclipsing binary star composed by a $\sim$G4 IV subgiant star eclipsed by a late-type M8-L3 dwarf star. The size of the three components is scaled to their actual sizes as determined in this paper.}
\label{fig:schematic}
\end{figure}

Given these numbers, we conclude that the most plausible scenario for the \koi{} system is that both components \kA{} and \kB{} are bound (and hence are coeval), where \kA{} is a giant star showing solar-like pulsations (Sect.~\ref{sec:asteroseismology}) and \kB{} is a subgiant star hosting \kC{}. {The bound scenario is also reinforced by the very similar systemic RV of both components around $-25$ km/s as shown in Fig.~\ref{fig:2comp}}. Figure~\ref{fig:schematic} shows a schematic view of the system given the above-mentioned architecture and including the information on the individual components derived in the subsequent sections.

\subsection{Spectroscopic analysis}
\label{sec:spec_analysis}

Among the two bright stars in the \koi{} system, the sharp component from the CCF analysis (corresponding to \kA as demonstrated in \ref{sec:who}) is the one that dominates the spectrum owing to its largest contrast and narrower spectral lines. Hence, in this section we consider only this source (star A) and neglect the possible effects in the spectrum due to the shallow component (star B). 

We derived the stellar atmospheric parameters (T$_{\rm eff}$, $\log{g}$, micro-turbulence, [Fe/H]) and respective uncertainties using the \texttt{ARES} and \texttt{MOOG} codes, following the same methodology described in \cite{sousa14} and \cite{santos13}. The equivalent widths (EW) of iron lines were measured on the combined CAFE spectrum of KOI3886 using the \texttt{ARES} code\footnote{The last version of \texttt{ARES} code (\texttt{ARES} v2) can be downloaded at \url{http://www.astro.up.pt/$\sim$sousasag/ares}} \citep{sousa07,sousa15}. We used a minimization process to find ionization and excitation equilibrium and converged to the best set of spectroscopic parameters. This process makes use of a grid of Kurucz model atmospheres \citep{Kurucz-93} and the radiative transfer code \texttt{MOOG} \citep{Sneden-73}. Since the star is cooler than 5200~K we used the appropriate iron line list for our method presented in \citet[][]{Tsantaki-2013}. The results of this analysis are shown in the upper part of Table~\ref{tab:starA} and point to the spectrum dominant star (\kA{}) to be an early-K type star ascending the RGB, hence being the source of the solar-like oscillations seen in the \textit{Kepler} light curve. 

\begin{table}[]
\setlength{\extrarowheight}{3pt}
\caption{\label{tab:starA} Spectroscopic (Sect.~\ref{sec:spec_analysis}) and asteroseismology (Sect.~\ref{sec:asteroseismology}) parameters of \kA{}.}
\begin{tabular}{lc}
 
 \hline 
Parameter & \kA{}  \\ 
\hline 
\textit{Spectroscopy} \\
\hline
Effective temperature, $T_{\rm eff,A}$ & $4720\pm120$~K\\
Surface gravity, $\log{g}_{\rm A}^{\rm spec}$ & $2.54\pm0.24$ \\
Turbulent velocity, $v_{\rm turb}$ & $1.172\pm0.095$~km/s \\
Metallicity, [Fe/H] & $0.143 \pm0.065$ \\
 \hline
\textit{Asteroseismology} \\
\hline
Stellar mass, $M_{\rm \star,A}$ & $1.69^{+0.14}_{-0.10}~M_{\odot}$ \\
Stellar radius, $R_{\rm \star,A}$ & $11.13^{+0.32}_{-0.26}~R_{\odot}$ \\
Stellar luminosity, $L_{\rm \star,A}$ & $55.1^{+5.2}_{-2.4}~L_{\odot}$ \\
Surface gravity, $\log{g}_{\rm \star,A}^{\rm ast}$ &$2.577^{+0.011}_{-0.009}$ \\
Stellar age & $2.05^{+0.56}_{-0.28}$~Gyr \\
 \hline
 
 \end{tabular}

\end{table}

\subsection{Asteroseismology}
\label{sec:asteroseismology}
\kA{} was previously classified as a first-ascent red giant by several authors \citep{elsworth17,hon17,vrard18} based on the analysis of its oscillation spectrum. The star physical parameters were previously determined through scaling relations to be $M_{\star,A} = 1.81\pm0.11$~\Msun{} and $R_{\star,A} = 11.61\pm0.25$~\Rsun{} \citep{yu18}. In this work, we make use of this prior information regarding the {evolutionary} state of the star to conduct a detailed asteroseismic analysis and determine precise stellar fundamental properties (full details of the analysis will be presented in a follow-up paper; T. Campante et al. 2021, in preparation).

We based our analysis on a custom \textit{Kepler} light curve generated by the KADACS pipeline (\textit{Kepler} Asteroseismic Data Analysis and Calibration Software; \citealt{garcia11,garcia14}) spanning nearly four years. Figure~\ref{fig:background} shows the corresponding power spectrum, which reveals a clear power excess due to solar-like oscillations at $\sim50\:\mu\rm{Hz}$. We measured the large frequency separation, $\Delta\nu$, and the frequency of maximum oscillation amplitude, $\nu_{\rm max}$, using well-tested automated methods \citep{campante17,campante19,corsaro20}, which returned $\Delta\nu = 4.60 \pm 0.20 \: \mu \rm{Hz}$ and $\nu_{\rm max} = 46.9 \pm 0.3 \: \mu\rm{Hz}$. We next extracted individual mode frequencies from the power spectrum using the FAMED pipeline (Fast and AutoMated pEak bagging with DIAMONDS; \citealt{corsaro20}). A total of 34 modes of angular degree $\ell$ = 0, 1, and 2 (including a number of mixed $\ell=1$ modes) were extracted across seven radial orders (see Fig.~\ref{fig:echelle}).

A grid-based inference procedure was used to determine the fundamental properties of \kA{}. It consisted in fitting the individual oscillation frequencies along with the classical constraints $T_{\rm eff}$, $\log{g}$, and [Fe/H], following the approach described in \cite{li20}, without considering interpolation and setting the model systematic uncertainty to zero. The grid of stellar models was computed with recourse to the Modules for Experiments in Stellar Astrophysics ({\sc mesa}, version 12115) \citep{paxtonetal2011,paxtonetal2013, paxtonetal2015,paxtonetal2018} and the corresponding adiabatic model frequencies with \textsc{GYRE} (version 5.1; \citealt{townsend13}). The microscopic and macroscopic physics adopted in the construction of the grid followed closely that described in \cite{li20}, but a different solar chemical mixture was considered, namely, [$(Z/X)_{\odot}$ = 0.0181] provided by \citet{asplund09}. Models in the grid varied in stellar mass ($M_{\star}$) within 0.8 – 2.2~\Msun{} in steps of 0.02~\Msun{}, initial helium fraction ($Y_{\rm init}$) within 0.24 – 0.32 in steps of 0.02, initial metallicity ([Fe/H]) within -0.5 – 0.5 in steps of 0.1. Moreover, four values were considered for the mixing length parameter associated with the description of convection, namely,  $\alpha_{\rm MLT}$ = 1.7, 1.9, 2.1, and 2.3. We considered convective overshooting in the core and applied the exponential scheme given by \citet{herwig00}. We set up the overshooting parameter as a function of the mass $f_{{\rm ov}}$ = (0.13$M$ - 0.098)/9.0 and used a fixed $f_{{\rm ov}}$ of 0.018 for models with a mass above $2.0$~\Msun{}, following the mass-overshooting relation found by \citet{2010ApJ...718.1378M}.

The stellar properties inferred from our modeling procedure are shown at the bottom part of Table~\ref{tab:starA}. These were obtained from the corresponding probability density distributions, where the value corresponds to the median and the uncertainties to the 16th and 84th percentiles, respectively. The frequencies of a representative best-fitting model are shown in the \'echelle diagram in Fig.~\ref{fig:echelle} for comparison with the observed frequencies. 

We should note at this point that no other seismic signal is seen in the power spectrum. This null detection is consistent with the stellar parameters obtained for \kB{} in Sect.~\ref{sec:system}. By using the relation $\nu_{\rm max} \propto g*T_{\rm eff}^{-1/2}$ scaled by solar values \citep{brown91,belkacem11}, this gives an expected $\nu_{\rm max}$ for \kB{} of $\sim$370 $\mu$Hz. This is above the Nyquist frequency of \textit{Kepler}'s 30-min cadence data of 280 $\mu$Hz (see Fig.~\ref{fig:background}). Even with a faster cadence (and hence higher Nyquist frequency), the wash-out (dilution) from the primary would make oscillations in \kB{} likely non-detectable.

\subsection{Global stellar parameters}
\label{Sect:PARAMS}

The procedure to compute the whole set of parameters for stars A and B, and object C requires feedback between (1) the combined spectroscopic and asteroseismology analysis (Sects.~\ref{sec:spec_analysis} and \ref{sec:asteroseismology}), (2) the SED fitting (Sect.~\ref{sec:SED}), (3) the analysis of the light curve (Sect.~\ref{sec:joint}), and (4) the spectral fitting (Sect.~\ref{sec:SPEC}). Our analysis consisted of the following steps:
i) The parameters of star A $(T_{\rm eff,A},\log g_{\star,A})$ derived from
  (1) are used as constraints in (2) to estimate $T_{\rm eff,B}$; (2)
  is not very sensitive to stellar gravity and metallicity, therefore
  these parameters cannot be obtained in the first step.
ii) The age of the system estimated from (1), and the value of
  $T_{\rm eff,B}$, put limits in the Hertzprung-Russell (HR) diagram $\log g_{\star} - T_{\rm eff}$
  to the position of star B, which is used to set a constraint to
  the priors for the mass and radius of this star to be used in (3).
iii) The analysis in (3) provides constraints on $(M_{\star,B},R_{\star,B})$ and $(M_{\star,C},R_{\star,C})$. The values for star B are used to refine the results from (2).
iv) The position of star A in the HR diagram is refined by assuming that both A and B are coeval, and hence both should belong to the same isochrone. This allows us to recompute
  $(M_{\star,A},R_{\star,A})$.
v) Using the values of $(T_{\rm eff},\log g_{\star})$ for stars A and B, and
  the contrast at every wavelength from (2), a combined model spectrum
  for A and B are built and compared with the observed spectrum. This is
  a mandatory control to ensure the quality of the fit.
vi) The results $(T_{\rm eff},R_{\star})$ for A and B can be used to compute the
  correponding luminosities. Making use of the integration of the models
  obtained in (2), an estimate of the distance to both stars --a parameter
  that {\it never} enters the analysis-- is feasible.
  Since the working hypothesis is that the system is bound, slight
  refinements of the parameters are carried out until the values of
  the distances are the same, within the uncertainties, and the whole
  set of parameters is self-consistent.

In the following subsections, we give some specific details of the procedures (2) SED fitting and (4) spectral fitting.

\subsubsection{SED fitting}
\label{sec:SED}

The photometry in Table \ref{Table:PHOT} corresponds to the combined light of stars A and B, plus the eclipsing object C. Therefore we attempted to carry out a two-model fit fulfilling some observational constraints\footnote{The contribution of object C to the total flux at   any wavelength is negligible.}. Since data in the $i'K_{\rm s}$ bands (corresponding to the combined light of stars A and B) are available (Table \ref{Table:PHOT}), and the contrasts between the fluxes of both objects in these bands are known from Robo-AO ($i'$, 1.13 mag, \citealt{ziegler17}) and PHARO ($K_{\rm s}$, 2.23 mag, \citealt{Furlan17}\footnote{\url{https://vizier.u-strasbg.fr/viz-bin/VizieR-3?-source=J/AJ/153/71/table8}}), the first constraint to build a composite model is that the individual models reproduce the fluxes at $(i'K_{\rm s})$(A) and $(i'K_{\rm s})$(B). A second constraint is that the contribution of the model of star B compensates the deficit of flux at wavelengths shorter than 5000~\AA\ and, in particular, the NUV point that star A cannot account for.  The third constraint is that during the fitting procedure, the models for both stars are reddened with the same value of the color excess $E(B\!-\!V)$ since we work under the hypothesis that the system is bound (see Sect.~\ref{sec:who}); that is, both stars are at the same distance, so they must undergo the same interstellar absorption. Synthetic Castelli-Kurucz {\sc atlas9} models \citep{Castelli03} are used throughout.

\begin{figure}
\centering
\includegraphics[width=0.5\textwidth]{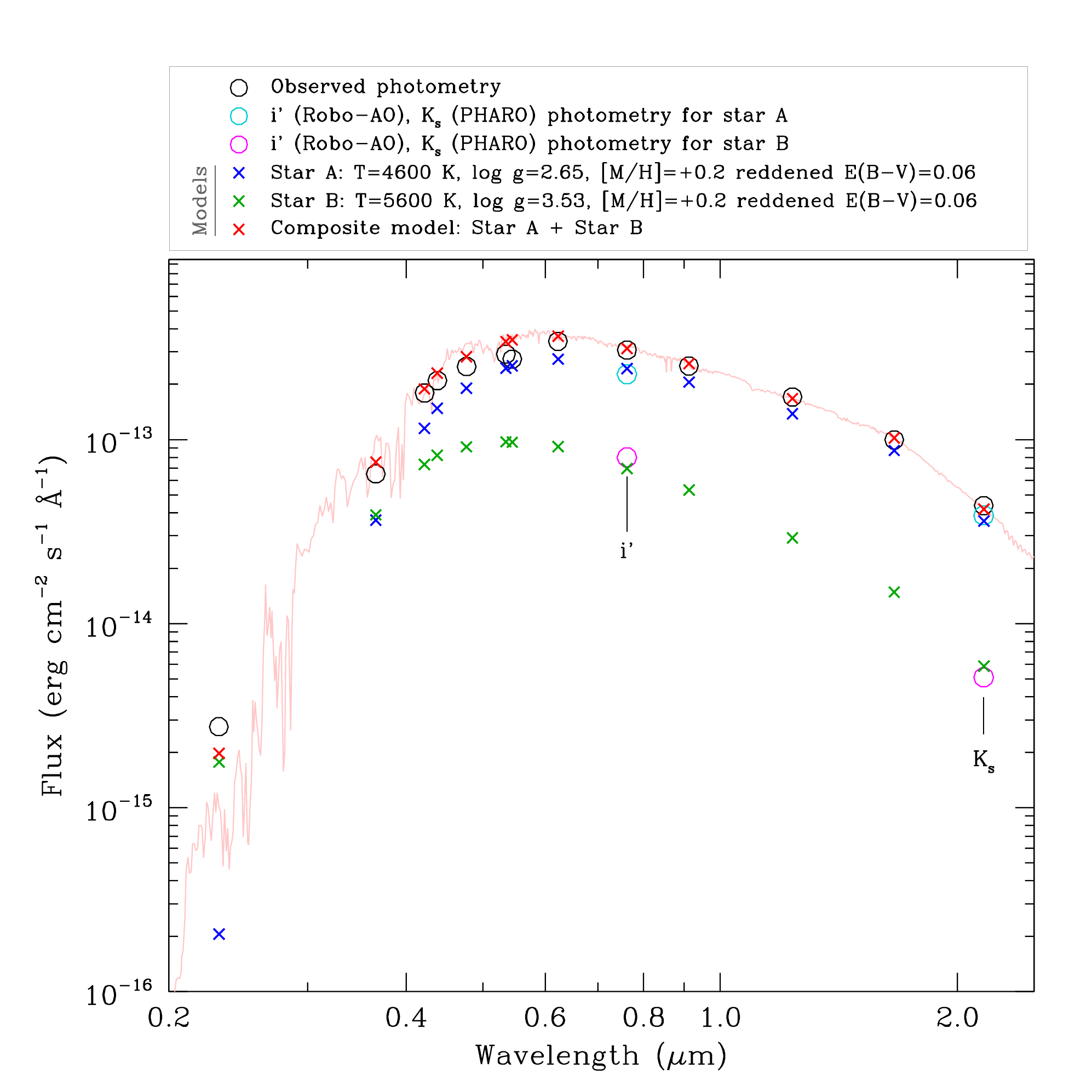}
\caption{Final result of the iterative process fitting the SED. The observed photometry is plotted as empty black circles, empty cyan and purple circles represent the fluxes at $i'K_{\rm s}$ for stars A and B, respectively. The synthetic photometry on the models for stars A and B and the composite --final-- model are plotted as blue, green, and red crosses. The composite model is also plotted in pale red.}
\label{fig:SED}
\end{figure}

\begin{table}[!htb]
\caption{Parameters of stars A and B as derived in Sect.~\ref{Sect:PARAMS}.}
\label{Table:PARAMS}
\begin{tabular}{ccccc}
  \hline\hline
  \noalign{\smallskip}
Star  & $T_{\rm eff}$ (K) &  $M_*/M_\odot$ & $R_*/R_\odot$ & $L_*/L_\odot$ \\
\noalign{\smallskip}
\hline
\noalign{\smallskip}
 A    & 4600$\pm$100  & 1.75$\pm$0.10 & 10.40$\pm$0.10 & 43.3$\pm$9.5 \\ 
 B    & 5600$\pm$100  & 1.61$\pm$0.05 &  3.61$\pm$0.09 & 11.5$\pm$1.0 \\
\noalign{\smallskip}\hline
\end{tabular}
\end{table}

\begin{figure}
\centering
\includegraphics[width=0.5\textwidth]{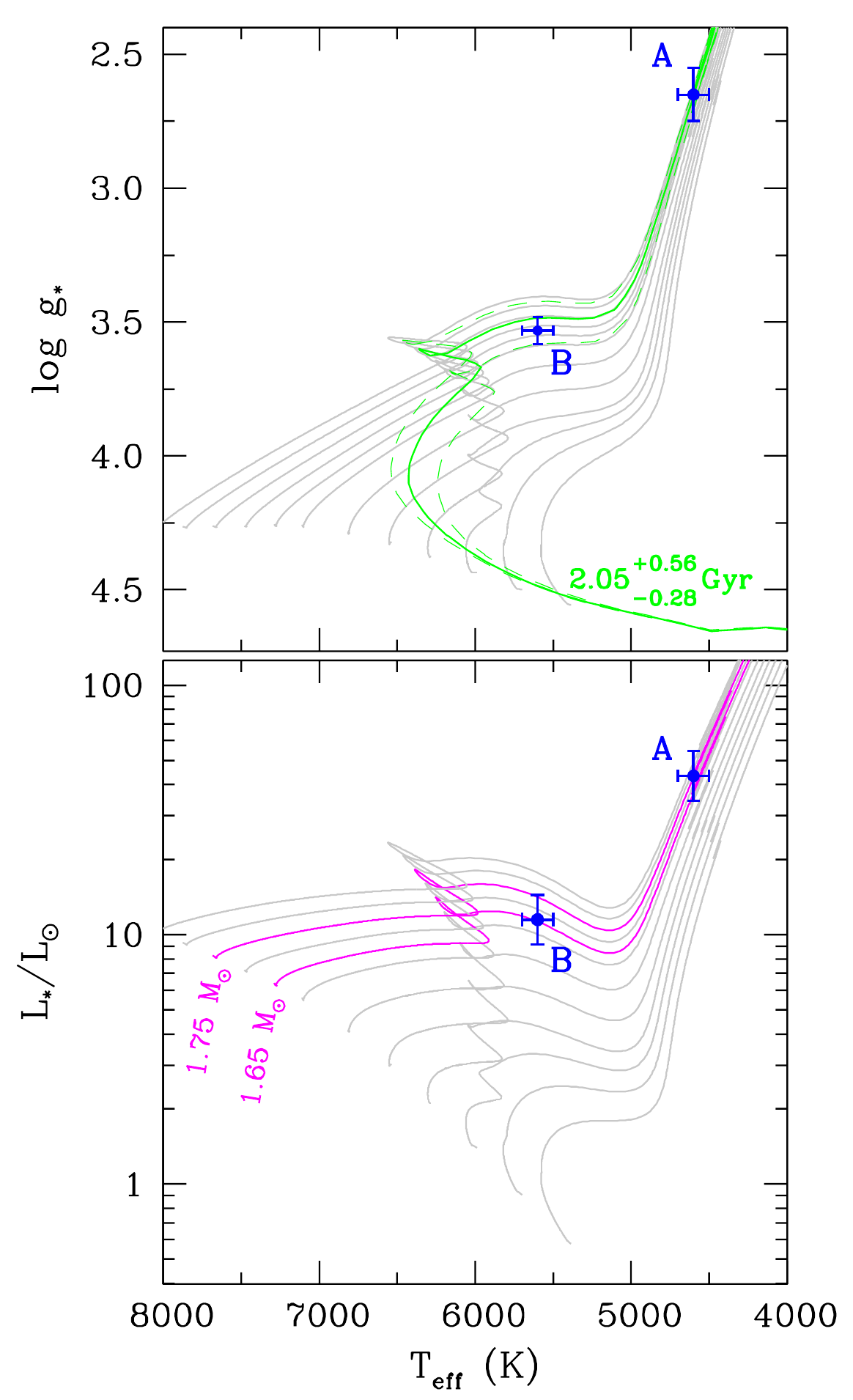}
\caption{Position of the stars in the HR diagram. PARSEC 2.1s tracks and isochrones with $Z\!=\!0.03$ have been used. The location  of the stars is consistent with the age found from the  asteroseismology analysis ($\sim\!2$ Gyr). Tracks of masses 1.0, 1.1,  1.2, 1.3, 1.4, 1.5, 1.6, 1.65, 1.70, 1.75, 1.8, and 1.85 $M_\odot$
  are plotted.}
\label{Fig:HR}
\end{figure}

\begin{figure*}
\centering
\includegraphics[width=0.9\textwidth]{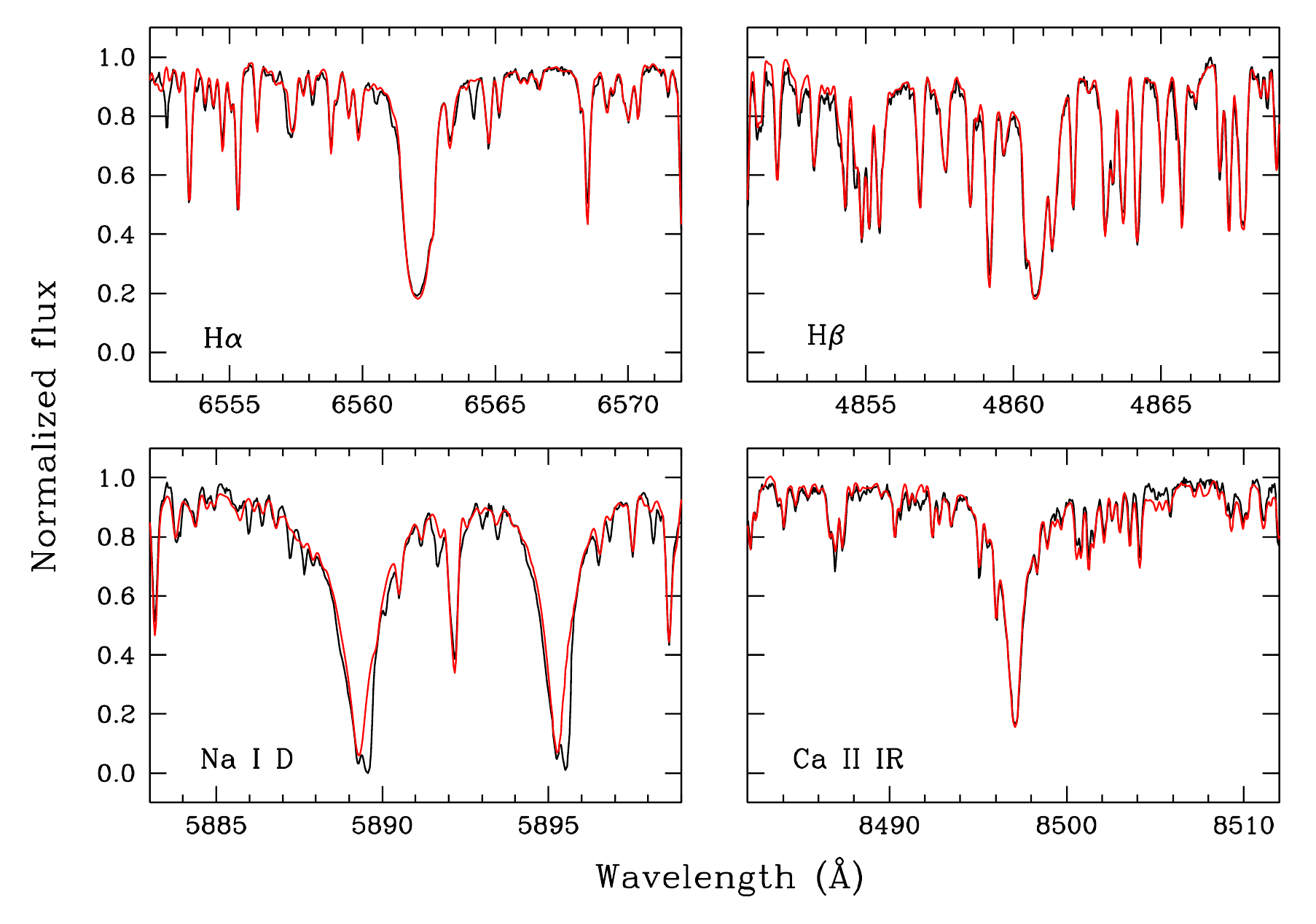}
\caption{CAFE spectrum of KOI-3886 (black) and the composite model
  built with the template spectra of real stars with types K2 III and G5 IV
  (red). Regions around H$\alpha$, H$\beta$, Na {\sc ii} D, and the
  Ca {\sc ii} 8498 \AA{} of the IR triplet lines are plotted.}
\label{Fig:SPEC}
\end{figure*}

Table \ref{Table:PARAMS} and Fig. \ref{fig:SED} show the final result of the iterative process fitting the SED. The observed photometry is plotted as empty black circles, empty cyan and purple circles represent the fluxes at $i'K_{\rm s}$ for stars A and B, respectively, computed from the contrasts provided by the high-resolution imaging observations; the synthetic photometry on the models for stars A and B, both of which are reddened with $E(B\!-\!V)\!=\!+0.06,$ are plotted as blue and green crosses, respectively, and the composite final model is plotted as red crosses and as a pale red line\footnote{ The GALEX NUV filter is so broad (FWHM$\simeq$800 \AA), and the SED so steep in this region that the point representing the synthetic photometry at this band,    plotted at the effective wavelength of this filter does not lie on the low-tesolution model itself}. It is apparent that the final model, although approaching the observed GALEX NUV flux, does not  quite reach that value; forcing the model to do so would push upward the fluxes at wavelengths around 4000 \AA, worsening the overall quality of the fit. A possible reason for the deficit of NUV flux could be the fact that the models used are purely photospheric; that is, they do not take into account any effect in the UV continuum from the presence of a chromosphere. \citet{Franchini98} carried out a comparison between observed and computed UV SEDs for a sample of  53 field G-type stars, concluding that UV excess shortward of 2000  \AA{} was evident for all the program stars. Taking into account the star B is a G subgiant, the small remaining discrepancy between the observations and model could be explained in this context.

Figure \ref{Fig:HR} shows the position of stars A and B in the HR diagrams $\log g_{\star} - T_{\rm eff}$ and $L_{\star}/L_\odot - T_{\rm eff}$. PARSEC 2.1s tracks and isochrones\footnote{\url{https://people.sissa.it/~sbressan/parsec.html}} \citep{Bressan12,Bressan13} have been used. During the iteration procedure it was found that a metallicity $Z\!=\!0.03$ ([M/H]$\simeq$+0.2) was more appropriate than the solar one, which agrees with the estimate from the asteroseismology analysis of star A (Sect.~\ref{sec:asteroseismology}). In the top panel the isochrone $2.05^{+0.56}_{-0.28}$ Gyr from the asteroseismology analysis is plotted; the positions of the stars are totally consistent with that estimate. 

A test about the robustness of the results is the computation of the distance to both stars, a parameter that did not enter and was not imposed at any moment in the calculations. From the values of $T_{\rm eff}$ and $R_{\star}$ the stellar luminosity can be found using the well-known expression $L_{\star} = 4\pi \sigma R_{\star}^2 T_{\rm eff}^4$. On the other hand the SED fitting provides us with models for the two stars, scaled to the observed photometry. Therefore the luminosity can also be written in terms of the integral of the dereddened model, $F_{\rm obs}$, as $L_{\star} = 4\pi d^2 F_{\rm obs}$; in both expressions of $L_{\star}$ the only unknown is the distance, $d$. Using the corresponding numbers for each star we obtain $d_{\rm A}$=685$\pm$75 pc and  $d_{\rm B}$=662$\pm$29 pc, that is, totally consistent with both stars being at the same distance. And interestingly, this is consistent within the uncertainties with the \textit{Gaia} distance to the two other sources in the field described in Sect.~\ref{sec:gaia}, thus suggesting a potential even more complex multi-stellar system.

\subsubsection{Spectral fitting}
\label{sec:SPEC}

In this section we show the results of the comparison of the CAFE spectrum with a combination of spectra of real stars taken as templates. For this exercise, we used spectra of HD 140573 (K2 III) (star A) and HD 152311 (G5 IV) (star B).  These spectra were downloaded from the ESO website\footnote{Field Stars Across H-R Diagram:\\ \url{http://www.eso.org/sci/observing/tools/uvespop/field\_stars\_uptonow.html}}. The data in this collection were obtained with UVES/VLT, therefore the template spectra were mapped onto the resolution of the CAFE spectrograph.

Figure \ref{Fig:SPEC} shows the comparison of the CAFE spectrum and the composite spectrum from the templates. No relative shift in RVs was applied when combining the spectra of the two stars. To carry out the combination, the template spectra of stars A and B were weighted at each wavelength according to the contrasts provided by the SED fitting.

The agreement between the CAFE spectrum and the composite model is remarkable, the only exception being the absorption at the red side of the two lines of the Na {\sc i} doublet, at around +15 km/s, which is possibly caused by the interstellar medium.

\subsection{Joint photometric and radial velocity analysis}
\label{sec:joint}
Based on the information explained in the above sections, we performed a joint modeling of the \textit{Kepler} light curve and both HERMES and CAFE RV data extracted using the two-component analysis of the CCF explained in Sect.~\ref{sec:FP}. In order to avoid the large photometric variations due to the pulsations and granulation of the giant component \kA{}, we used the phase-folded and binned \textit{Kepler} light curve instead of the full dataset. In this case, we used the \texttt{ellc} package \citep{maxted16b} to model the eclipse and phase curve variations. Given the effective temperature derived from the SED analysis for \kB{}, we used a Gaussian prior for its mass centered at 1.6~\Msun{} with a width of 0.1~\Msun{} and truncated to ensure positive masses only. The priors on the radii of components B and C and the mass of component C were left uniform in a comprehensive range. We used Gaussian priors for the quadratic limb-darkening coefficients centered at the values estimated by the tables from \cite{claret11} for the \textit{Kepler} band in a trilinear interpolation with its effective temperature, surface gravity, and metallicity. The period and the mid-time of the primary eclipse were set to Gaussian distributions based on preliminary analysis of the light curve. Additionally, we included the flux ratio of components B to C as a free parameter with a flat prior on the whole possible physical range. Also, owing to the presence of \kA{} inside the \textit{Kepler} aperture, we added a dilution factor using the same definition as in \cite{juliet}. Based on our multicolor, high spatial resolution images, we can provide an accurate prior for this factor by measuring the contrast magnitude between \kA{} and \kB{} in the \textit{Kepler} band, which we estimate to be $\Delta m=0.9$~mag. This implies a dilution factor of $\mathcal{D}=0.28$. Finally, we added a flux level ($F_0$) and a jitter factor ($\sigma_{\rm Kepler}$) for the photometric time series and a systemic velocity ($\delta$) and jitter ($\sigma$) for each of the two RV instruments. The parameters and their priors are shown in Table~\ref{tab:joint}.

\begin{figure}
\centering
\includegraphics[width=0.48\textwidth]{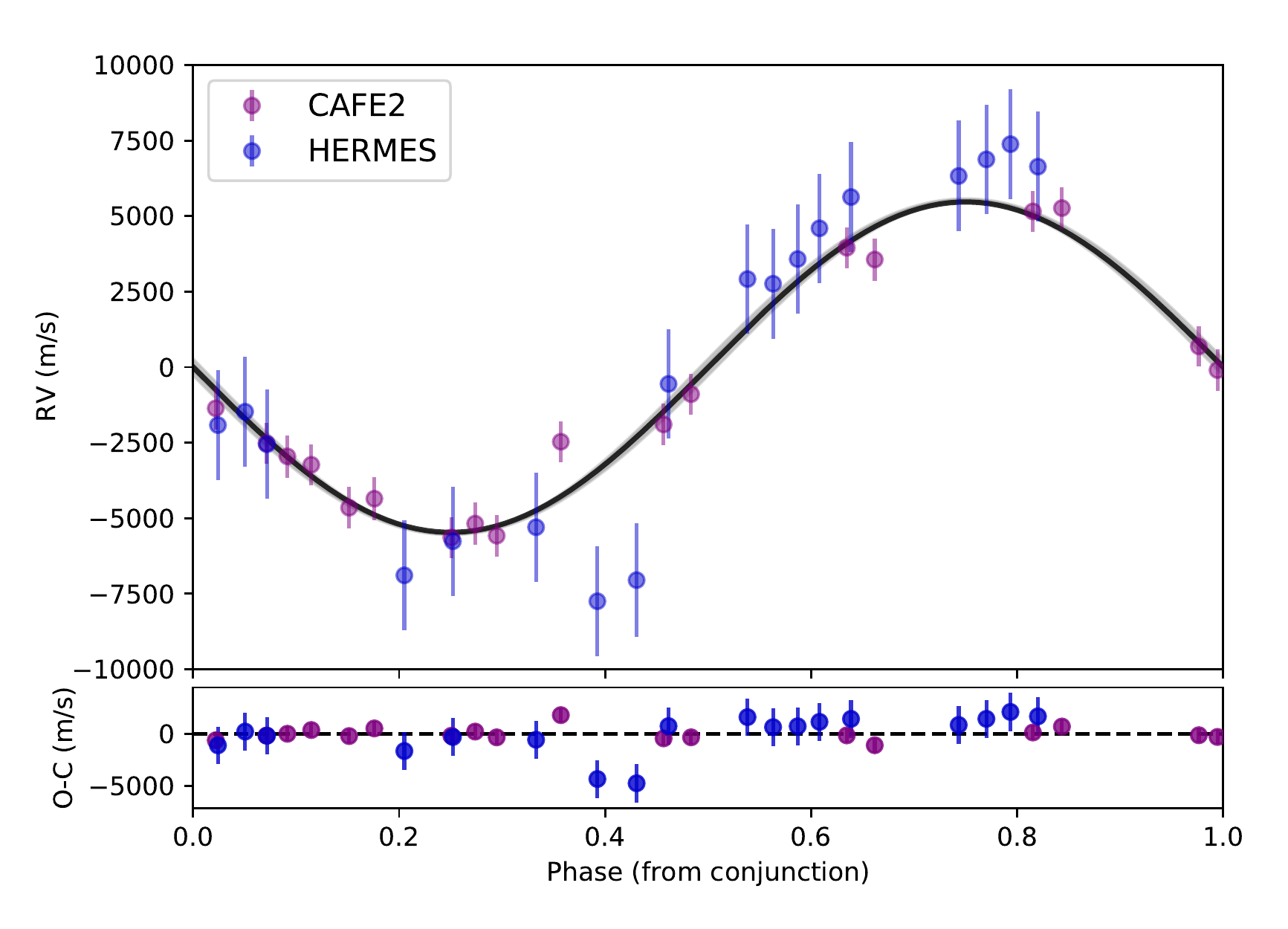}
\caption{Radial velocity of \kB{} variations phase-folded with the orbital period of \kC{}. The median model is shown as a solid black line and the corresponding 95\% confidence interval is shown as a shaded gray region.}
\label{fig:RV}
\end{figure}

We used a MCMC approach to model this dataset by using the \texttt{emcee} package. In our final run, we used 70 walkers (four times the number of parameters) and a first burn-in phase with 40\,000 steps followed by  a final production step with 20\,000 steps centered around the maximum probability parameter space found in the previous phase. This analysis leads to convergence of the chains. The marginalized posterior distribution of all parameters and their inter-dependences are shown in Fig.~\ref{fig:corner} and the median and confidence intervals are presented in Table~\ref{tab:joint}. The results show that \kC{} is an inflated brown dwarf ($M_{\star,C}=66.1^{+4.1}_{-3.2}$~\Mjup{}, $R_{\star,C}=1.524\pm0.070$~\Rjup{}) orbiting close to the subgiant star \kB{}. The median models for the light curve and RV signals are shown in Fig.~\ref{fig:LCphase} and Fig.~\ref{fig:RV}, respectively.

\section{Discussion}
\label{sec:discussion}

\subsection{Eclipsing brown dwarf inflated by its subgiant host}
Despite the initial planetary appearance of KOI-3886.01, the comprehensive analysis described in the previous sections shows a very different view of the overall system and, in particular, of the eclipsing object. The inferred mass of \kC{} consistently estimated by the RV and phase-curve modulation effects provides a value of $66.1\pm3.5$~\Mjup{}. This value clearly places \kC{} in the brown dwarf regime. 

Eclipsing binaries, specifically those with an independent age estimate, are true Rosetta stones for improving evolutionary models and understanding the properties of isolated stars, such as the characteristics of their atmospheres, their evolution, and dependence on stellar mass. \koi{} is a unique system for the rare combination of its members and configuration: a giant $\sim$K2 III star, essential to accurately determine the system age plus an eclipsing pair, composed of a subgiant $\sim$G4 IV star and a brown dwarf, the spectral type of the latter probably being in the L-T transition (i.e., effective temperature below 2250 K); see Fig.~\ref{fig:schematic}.

Several studies have found eclipsing binaries in stellar associations whose ages are known. These are the cases of M11, a $\sim$200 Myr old open cluster (see the case of KV29, a massive system \citealt{Bavarsad2016}); the Hyades, which is 625-790\,Myr, and several interesting systems (HD\,27130, which has a solar-like primary by \citealt{Brogaard2021}, and vA\,50, a M4 plus, perhaps, a planet \citealt{David2016b}); Praesepe,  which is 600-800 Myr (several M-type binaries; \citealt{Gillen2017}); several young stellar associations such as UpperSco, 5-10 Myr (USco J161630.68-251220.1, a M5.5 primary with a very low-mass star close to the substellar limit, \citealt{Lodieu2015}; or the transiting brown dwarf\,RIK\,72\,b, \citealt{David2019}), or the first ever discovered eclipsing binary brown dwarfs, which is located in the 1 Myr massive association in Orion (\citealt{Stassun2006}); and in globular clusters, such as NGC\,6362, a 11.67 Gyr association that includes V40, a solar-like pair (\citealt{Kaluzny2015}).

In the field, other systems containing an eclipsing binary, which are interesting for different reasons, have been analyzed by \cite{Jackman2019} or \cite{Lester2019a, Lester2019b, Lester2020}, just to provide a few examples. In particular, the combination of photometric, spectroscopic, and interferometric data (coming from the CHARA array, e.g.) can be used to provide a 3D orbit and greatly reduce the uncertainties of the system parameters. The advent of TESS has allowed detailed analysis of M plus brown dwarf pairs, such as those presented in \cite{Carmichael2019, Carmichael2020, Carmichael2021}. Of special interest is the triple system HIP\,96515 (\citealt{Huelamo2009}), which contains an eclipsing binary composed by M1 and M2 dwarfs orbiting each other every 2.2456~days, and a white dwarf located at $\sim$8 arcsec. In a sense, this system represents the future of \koi{} once the giant evolves and sheds its external layers.

\begin{figure*}
\centering
\includegraphics[width=0.48\textwidth]{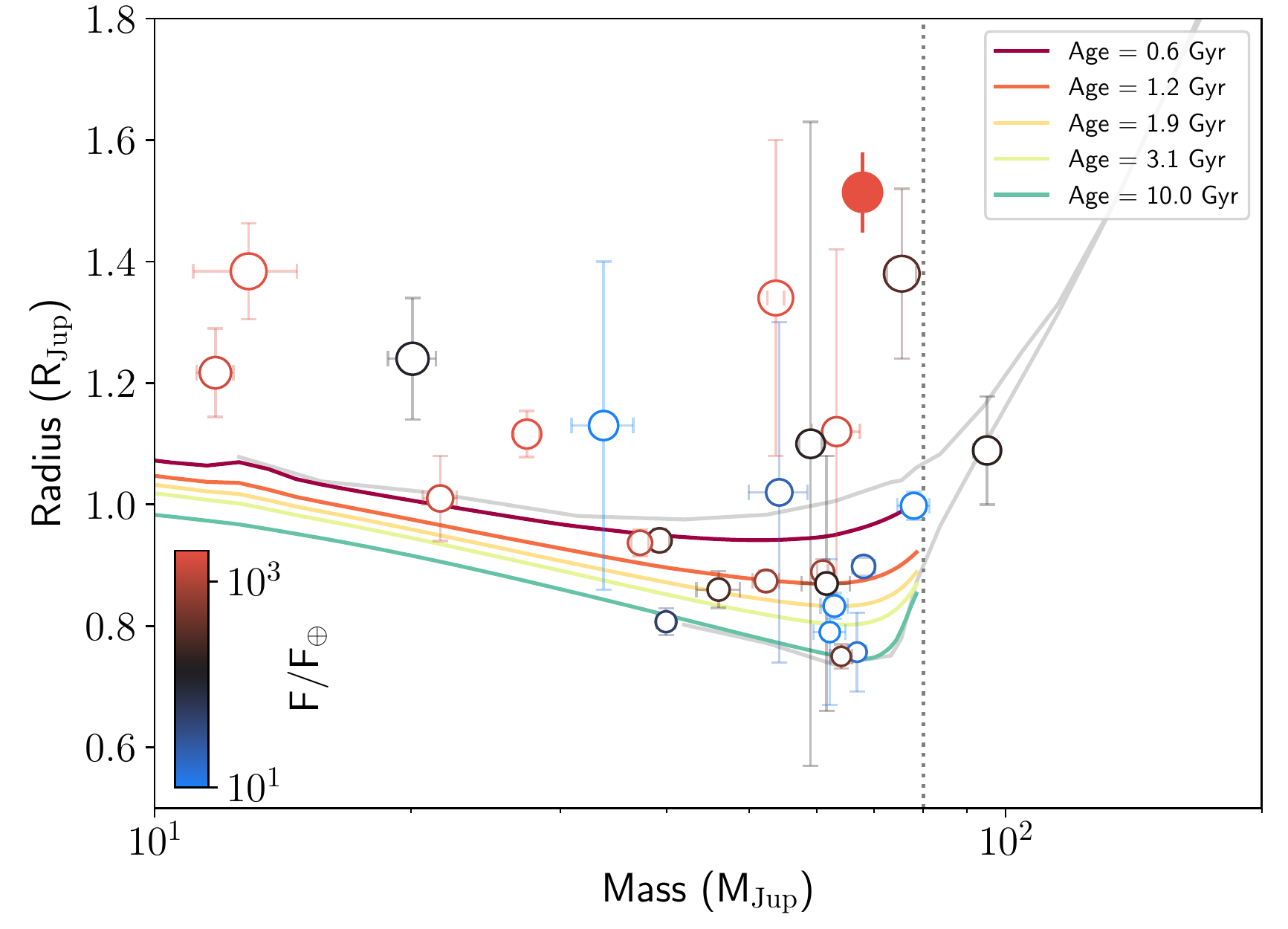}
\includegraphics[width=0.48\textwidth]{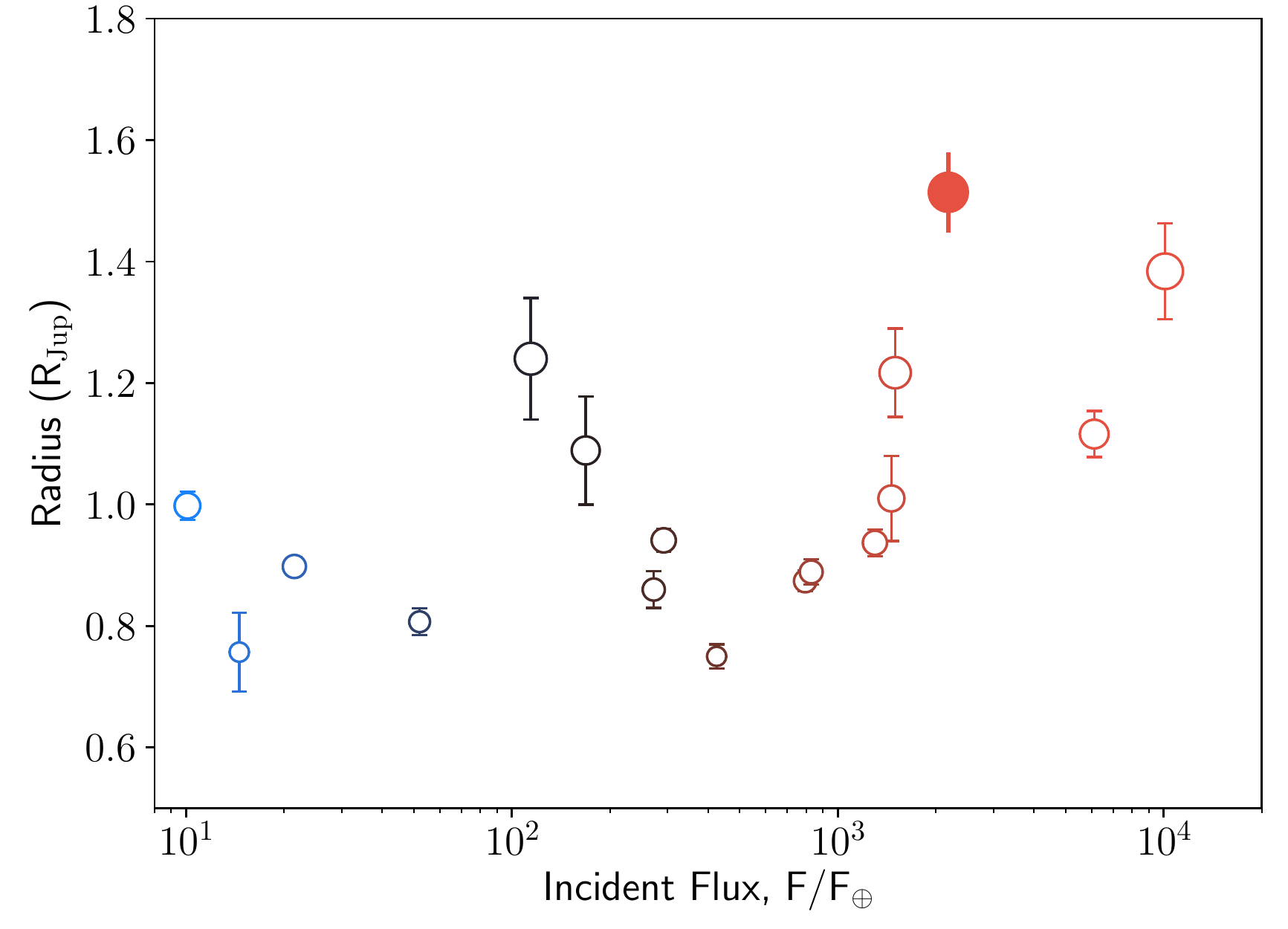}
\caption{Left panel: Mass-radius diagram in the brown dwarf regime for all known eclipsing brown dwarfs (open symbols) from \cite{carmichael20} (and references there in) and the latest discoveries of TOI-503 \citep{subjak20}, NGTS-7A \citep{jackman19}, and TOI-263 \citep{parviainen20,palle21}. \kC{} is represented by the filled symbol. The color code indicates the incident flux from their primary star in units of Earth's flux received from the Sun. Isochrones for different ages from \cite{phillips20} are shown in colors, while the stellar tracks from \cite{baraffe15} for 1 and 12 Gyr are shown in gray. The dashed vertical line marks the hydrogen burning limit between stars and brown dwarfs. Right panel: Radius vs. incident flux of the same sample of brown dwarfs as in the left panel, showing the increasing evidence of irradiation as the source of inflation for these objects. }
\label{fig:MR}
\end{figure*}

In a mass-radius diagram (see Fig.~\ref{fig:MR}, left panel), \kC{} appears as the most inflated brown dwarf known to date. The incident flux received from its host star (\kB{}) is about 2100 times that received by the Earth from the Sun, and is one of the most irradiated brown dwarfs known (see Fig.~\ref{fig:MR}, right panel). This irradiation might well be the cause of the atmospheric inflation of this massive brown dwarf. \kC{} is also the first brown dwarf known to eclipse an evolved star in the subgiant phase. The presence of the two evolved stars in the system, also makes \koi{} a dynamically and evolutionary interesting system to follow up.  

\subsection{Future evolution of the \koi{} system}

We first consider the implications of the evolution of \kB{} in the orbit of the substellar object, \kC{}. In the current orbit, \kC{} is unstable to tidal dissipation, so it will eventually experience orbital decay and end up inside the envelope of the subgiant. We estimated that the shrinking of the orbit from tidal decay takes 340~Myr from now, assuming a standard mass-loss rate for the star and typical tidal parameters. This is the time it takes to the star to evolve from the observed radius of 3.6~\Rsun{} to a stellar radius with a value of 8.9~\Rsun{}. In its current state, we estimate the shrinking rate to be $7.4\times10^{-9}$~au/yr, which translates into a period shrinking of 0.033 seconds/yr, which is difficult to measure in a decade time span.  

Another issue altogether is the stability of C given that both B and C are in a binary system with the A component. The C component is in what is traditionally defined as a satellite-type or S-type orbit \citep{Dvorak86}, where the less massive body is orbiting the secondary star and the primary star can be considered as a perturber.  From the empirical expressions of \cite{Holman99}, we estimate the stable region around the subgiant for a circular binary orbit. We find the critical semimajor axis to be $a_{\rm C}^{\rm crit}=27.35$~au, which is inside the orbital distance determined for the substellar object around the star B ($a_{\rm C}=0.159$~au).

The evolution of the binary is expected to modify the size of this stable region, and stellar evolution can trigger dynamical instabilities that drive planets into chaotic orbits. As the stars loose mass via winds, the mass ratio of the binary changes and with it the critical distance for stability \citep{kratter12}. However, such a process does not happen in this system before C undergoes the orbital evolution described before. Although substantial mass-loss is responsible for a significant change in the critical stability radius, it does not occur until the star ascends the asymptotic giant branch after the red and horizontal branches.  At that time, the substellar object is not expected to be orbiting the secondary, since as calculated before, it is engulfed well before the star B reaches the tip of the RGB.

\section{Conclusions}
\label{sec:conclusions}

In this paper we have unveiled the nature of the \koi{} system and planet candidate. Thanks to a large number of follow-up observations and analysis {(mainly driven by the chromatic RV analysis and the use of different spectral masks to extract the RV; see Sect.~\ref{sec:FP})}, we could conclude that KOI-3886.01 is actually an eclipsing close-in brown dwarf (\kC{}) orbiting a $\sim$G4 star in the subgiant phase (\kB{}); both of these are bound to a $\sim$K2 star ascending the RGB (\kA{}). The eclipsing binary system has a projected separation to the RGB star of around 270 au. This relatively large separation implies an independent evolution of the binary system, not affected by the evolution and mass loss of the RGB star.

We inferred the properties of \kA{} through a combination of spectroscopic, asteroseismic, and SED analysis, and conclude that this star has a radius of 10.4~\Rsun{}, and an effective temperature of 4600~K and well-defined age of 2.05~Gyr. This star is the brightest of the two components seen in the high-spatial resolution images and is the dominant source in the optical spectrum of HERMES and CAFE data, corresponding to the sharp and narrow component of the CCF. Star \kB{} on its side corresponds to the slightly fainter source in the high-resolution images and contributes to the spectrum with broader and shallower spectral lines. This is reflected in the CCF where this component shows large periodic RV variations coincident with the \textit{Kepler} light curve eclipses. Through a thorough SED analysis, we could conclude that \kB{} should be bounded to \kA{} and, consequently, it must be in the subgiant phase with the same age. We infer an effective temperature of 5600~K for \kB{}, making the secondary a $\sim$G4~IV star. The bound nature of the two evolved stars puts important constraints on the mass of \kB,{} which become critical in estimating the mass of the eclipsing body. We used the RVs inferred from the HERMES and CAFE data and the \textit{Kepler} light curve to perform a joint analysis, in which we were able to conclude that \kB{} has a mass of 1.6~\Msun{} and a radius of 3.61~\Rsun{}, which agrees with its evolutionary stage. More importantly, we derive a mass of 66.1~\Mjup{} and a radius of 1.52~\Rjup{} to the third component in this system (\kC{}), placing it in the brown dwarf regime. The inferred radius is about 50\% larger than what theoretical models predict for this age. This inflation might be due to the strong stellar irradiation from \kB{} given the close-in orbit of this system, which has an orbital period of just 5.6 days around a subgiant star.

The dynamical analysis of this system indicates that the eclipsing binary is stable for the next 340 Myr. After that, the evolution of \kB{} will rapidly end in the disruption of \kC{} well before the B component reaches the tip of the RGB. The evolution of \kA{} will not affect the dynamical stability of the binary at any stage before it reaches the tip of the AGB, which will happen way after the brown dwarf has been engulfed by its host.

\koi{} fully complements the parameter spaces left unpopulated by other systems and can be considered as a stepping stone, which will help improve the models. \kC{} is the largest brown dwarf known to date and the first one known to eclipse an evolved star.

\begin{acknowledgements}
The authors thank the anonymous referee for taking the time to review this manuscript {and the language editor for a thorough spelling and grammatical review}.
J.L-B. acknowledges financial support received from ”la Caixa” Foundation (ID 100010434) and from the European Union’s Horizon 2020 research and innovation programme under the Marie Skłodowska-Curie grant agreement No 847648, with fellowship code LCF/BQ/PI20/11760023. This research has also been partly funded by the Spanish State Research Agency (AEI) Projects No.ESP2017-87676-C5-1-R and No. MDM-2017-0737 Unidad de Excelencia "Mar\'ia de Maeztu"- Centro de Astrobiolog\'ia (INTA-CSIC). 
The work by B.M. and E.V. is partially funded by the Spanish "Ministerio de Ciencia, Innovación y Universidades" through the national project "On the Rocks II" (PGC2018-101950-B-100; PI E. Villaver). 
This work was partially supported by FCT/MCTES through the research grants UIDB/04434/2020, UIDP/04434/2020 and PTDC/FIS-AST/30389/2017, and by FEDER - Fundo Europeu de Desenvolvimento Regional through COMPETE2020 - Programa Operacional Competitividade e Internacionaliza\c{c}\~ao (grant: POCI-01-0145-FEDER-030389). M.S.C. is supported by national funds through FCT in the form of a work contract. 
This work was supported by FCT - Funda\c{c}\~ao para a Ci\^encia e a Tecnologia through national funds and by FEDER through COMPETE2020 - Programa Operacional Competitividade e Internacionaliza\c{c}\~ao by these grants: UID/FIS/04434/2019; UIDB/04434/2020; UIDP/04434/2020; PTDC/FIS-AST/32113/2017 \& POCI-01-0145-FEDER-032113; PTDC/FIS-AST/28953/2017 \& POCI-01-0145-FEDER-028953; PTDC/FIS-AST/28987/2017 \& POCI-01-0145-FEDER-028987. 
T.C.~is supported by Funda\c c\~ao para a Ci\^encia e a Tecnologia (FCT) in the form of a work contract (CEECIND/00476/2018). 
A.A. acknowledges support from Government of Comunidad Aut\'onoma de Madrid (Spain) through postdoctoral grant `Atracci\'on de Talento Investigador' 2018-T2/TIC-11697
T.L. acknowledges the funding from the European Research Council (ERC) under the European Union‚ Horizon 2020 research and innovation programme (CartographY GA. 804752).
The research leading to these results has (partially) received funding from the European Research Council (ERC) under the European Union’s Horizon 2020 research and innovation programme (grant agreement N$^\circ$670519: MAMSIE) and from the KU~Leuven Research Council (grant C16/18/005: PARADISE). 
Based on observations obtained with the HERMES spectrograph, which is supported by the Research Foundation - Flanders (FWO), Belgium, the Research Council of KU Leuven, Belgium, the Fonds National de la Recherche Scientifique (F.R.S.-FNRS), Belgium, the Royal Observatory of Belgium, the Observatoire de Genève, Switzerland and the Thüringer Landessternwarte Tautenburg, Germany. 
This work has made use of data from the European Space Agency (ESA) mission {\it Gaia} (\url{https://www.cosmos.esa.int/gaia}), processed by the {\it Gaia} Data Processing and Analysis Consortium (DPAC, \url{https://www.cosmos.esa.int/web/gaia/dpac/consortium}). Funding for the DPAC has been provided by national institutions, in particular the institutions participating in the {\it Gaia} Multilateral Agreement.
We thank M. G\"unther for his help in using the \texttt{allesfitter} code \citep{allesfitter-paper,allesfitter-code}, although we finally did not use it for other reasons. 
This work has made use of the following python packages: 
\texttt{astropy} \citep{astropy:2013,astropy:2018}, 
\texttt{numpy} \citep{numpy}, 
\texttt{batman} \citep{kreidberg15}, 
\texttt{emcee} \citep{emcee}, 
\texttt{ellc} \citep{maxted16b}, 
\texttt{matplotlib} \citep{matplotlib}, 
\texttt{astroML} \citep{astroML}, and 
\texttt{corner} \citep{corner}.
\end{acknowledgements}

%
%


\bibliographystyle{aa} 
\bibliography{../../biblio2} 

\appendix

\section{Tables}

\begin{table*}[!htb]
\caption{Photometry of KOI-3886.}
\label{Table:PHOT}
\setlength{\tabcolsep}{10pt}
\begin{tabular}{rrcll}
  \hline\hline
  \noalign{\smallskip}
 \multicolumn{1}{c}{$\lambda$} & \multicolumn{1}{c}{Magnitude}   &   Flux  & Band & Source \\
 \multicolumn{1}{c}{(\AA)}     &                       & (erg cm$^{-2}$ s$^{-1}$ \AA$^{-1}$) & & \\
\noalign{\smallskip}
\hline
\noalign{\smallskip}
 2315.7 & 17.169$\pm$0.013  & 2.754(-15)$\pm$3.297(-17) & GALEX NUV$^\dagger$ & \cite{Beitia16}\\ 
 3660   & 12.016$\pm$0.022  & 6.520(-14)$\pm$1.321(-15) & Johnson $U$ & \cite{Bessell98}    \\
 4220   & 11.447$\pm$0.091  & 1.794(-13)$\pm$1.505(-14) & Tycho $B$  & \cite{Mann15}     \\
 4380   & 11.198$\pm$0.027  & 2.097(-13)$\pm$5.214(-15) & Johnson $B$ & \cite{Bessell98}     \\
 4770   & 10.706$\pm$0.020  & 2.497(-13)$\pm$4.599(-15) & Sloan $g'$ & \cite{Fukugita96,York00}   \\
 5350   & 10.352$\pm$0.054  & 2.914(-13)$\pm$1.450(-14) & Tycho $V$ & \cite{Mann15}      \\
 5450   & 10.306$\pm$0.050  & 2.739(-13)$\pm$1.262(-14) & Johnson $V$ & \cite{Bessell98}     \\
 6230   &  9.782$\pm$0.020  & 3.428(-13)$\pm$6.315(-15) & Sloan $r'$ & \cite{Fukugita96,York00} \\
 7620   &  9.465$\pm$0.020  & 3.069(-13)$\pm$5.653(-15) & Sloan $i'$ & \cite{Fukugita96,York00}   \\
 9130   &  9.288$\pm$0.020  & 2.516(-13)$\pm$4.636(-15) & Sloan $z'$ & \cite{Fukugita96,York00}  \\
12350   &  8.159$\pm$0.020  & 1.705(-13)$\pm$3.141(-15) & 2MASS $J$ & \cite{Cohen03}     \\
16620   &  7.638$\pm$0.020  & 9.978(-14)$\pm$1.838(-15) & 2MASS $H$ & \cite{Cohen03}       \\
21590   &  7.475$\pm$0.023  & 4.383(-14)$\pm$9.284(-16) & 2MASS $K_{\rm s}$ & \cite{Cohen03} \\
\noalign{\smallskip}
\hline
\noalign{\smallskip}
\multicolumn{4}{l}{$\dagger$ GALEX NUV magnitude is available from the Vizier catalog:}\\
\multicolumn{4}{l}{\url{https://vizier.u-strasbg.fr/viz-bin/VizieR?-source=J/ApJ/813/100}}
\end{tabular}
\end{table*}


\begin{table*}
\setlength{\extrarowheight}{3pt}
\caption{Derived stellar parameters from components B and C of the KOI-3886 hierarchical triple system through the joint (RV, eclipse, and phase modulations) analysis presented in Sect.~\ref{sec:joint}.}
\label{tab:joint}
\begin{tabular}{lll}
\hline\hline
Parameter & Priors & Posteriors \\
\hline
\textit{Orbital parameters} & & \\
\hline
Orbital period, $P_C$ [days] & $\mathcal{G}$(5.56648691,0.0001) & $5.566513^{+0.000043}_{-0.000043}$ \\
Time of mid-transit, $T_{\rm 0,C}-2400000$ [days] & $\mathcal{G}$(54966.03305,0.008) & $54966.0334^{+0.0012}_{-0.0012}$ \\
Orbital inclination, $i_{\rm C}$ [deg.] & $\mathcal{U}$(70.0,90.0) & $82.50^{+0.58}_{-0.53}$ \\
Orbit semimajor axis, $a_{C}$ [AU] & (derived) & $0.0720^{+0.0013}_{-0.0013}$ \\
Relative orbital separation, $a_{C}/R_{\star}$ & (derived) & $4.287^{+0.093}_{-0.085}$ \\
Incident flux, $F_{\rm inc,C}$ [$F_{{\rm inc},\oplus}$] & (derived) & $126.6^{+4.0}_{-2.6}$ \\

\hline
\textit{Star B} & & \\
\hline
Stellar mass, $M_{B}$ [$M_{\odot}$] & $\mathcal{T}$(1.6,0.1,1.4,1.8) & $1.608^{+0.087}_{-0.088}$ \\
Stellar radius, $R_{B}$ [$R_{\odot}$] & $\mathcal{T}$(3.5,0.3,2.5,4.5) & $3.61^{+0.11}_{-0.12}$ \\
Limb darkening coefficient, $u_1$ & $\mathcal{G}$(0.592,0.02) & $0.569^{+0.019}_{-0.020}$ \\
Limb darkening coefficient, $u_2$ & $\mathcal{G}$(0.141,0.02) & $0.131^{+0.020}_{-0.020}$ \\

\hline
\textit{Star C} & & \\
\hline
Stellar mass, $M_{C}$ [$M_{\rm Jup}$] & $\mathcal{U}$(0.0,55000.0) & $66.1^{+4.1}_{-3.2}$ \\
Stellar radius, $R_{C}$ [$R_{\rm Jup}$] & $\mathcal{U}$(0.0,100.0) & $1.524^{+0.070}_{-0.072}$ \\
Eclipse depth, $\Delta_{C}$ [ppt] & (derived) & $1.880^{+0.089}_{-0.088}$ \\
Eclipse duration, $T_{\rm 14,C}$ [hours] & (derived) & $8.873^{+0.071}_{-0.070}$ \\

\hline
\textit{Instrumental parameters} & & \\
\hline
LC level & $\mathcal{U}$(-500.0,500.0) & $58.9^{+6.7}_{-5.8}$ \\
Dilution factor & $\mathcal{T}$(0.28,0.01,0.2,0.4) & $0.2765^{+0.010}_{-0.0096}$ \\
LC jitter [ppm] & $\mathcal{U}$(0.0,2000.0) & $53.0^{+1.2}_{-1.1}$ \\
$\delta_{\rm HERMES}$ [km/s] & $\mathcal{U}$(-40.0,0.0) & $-25.86^{+0.42}_{-0.41}$ \\
$\sigma_{\rm HERMES}$ [km/s] & $\mathcal{U}$(0.0,2.1) & $1.81^{+0.20}_{-0.24}$ \\
$\delta_{\rm CAFE2}$ [km/s] & $\mathcal{U}$(-35.0,-10.0) & $-23.63^{+0.18}_{-0.17}$ \\
$\sigma_{\rm CAFE2}$ [km/s] & $\mathcal{U}$(0.0,2.0) & $0.66^{+0.17}_{-0.12}$ \\

\hline

\end{tabular}
\end{table*}


\section{Figures}

\begin{figure*}
\centering
\includegraphics[width=1\textwidth]{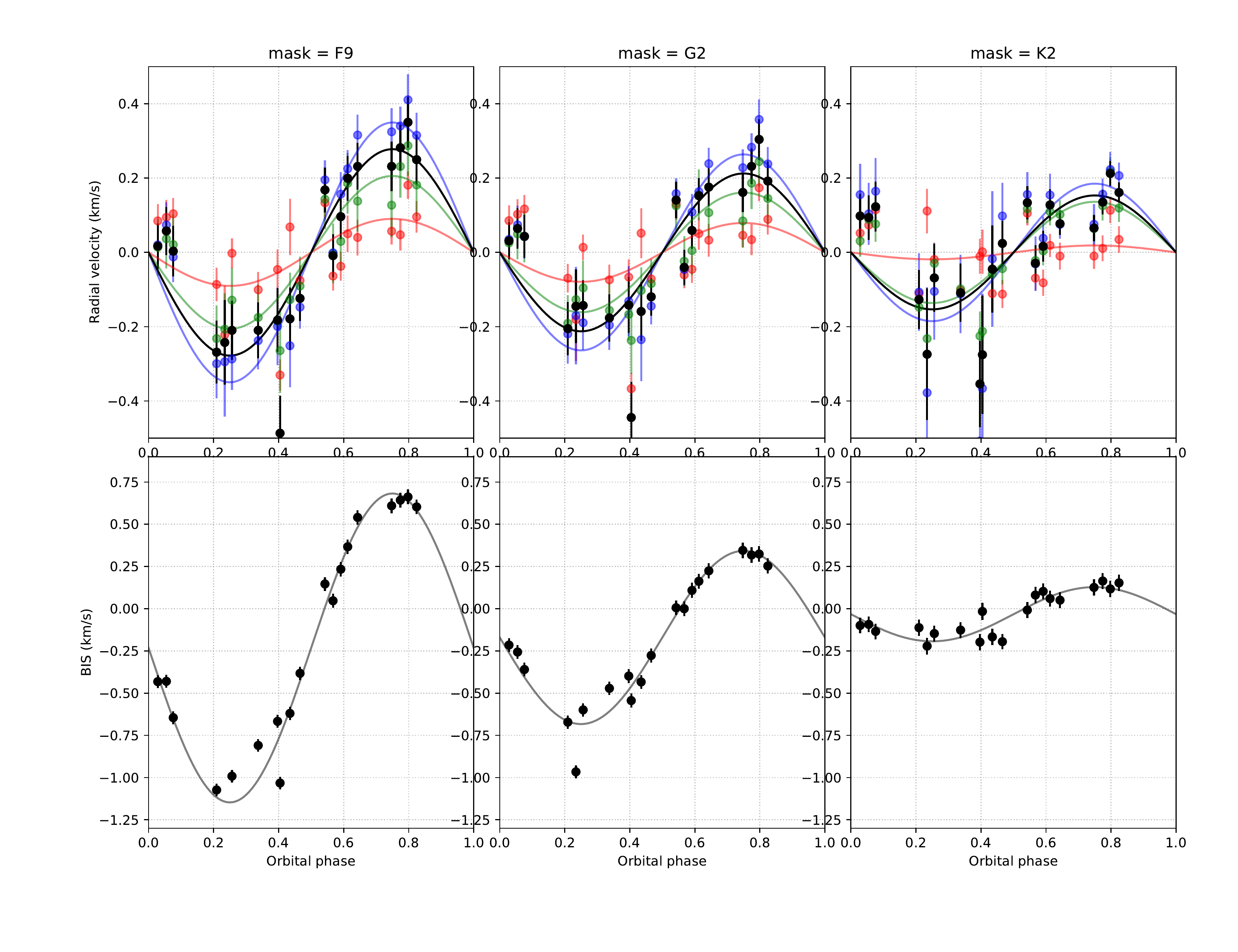}
\caption{Inferred RV (top panels) and bisector span (bottom panels) for different binary masks (F9, G2, and K2 from left to right panels). In the top panels, the different colors show the RVs estimated with three different wavelength ranges: blue (3800\AA-5100\AA), green (5100\AA-6400\AA), and red (6400\AA-7800\AA). A simple maximum-likelihood model fit to the corresponding dataset is shown with solid lines. All panels are phase-folded with the \textit{Kepler} eclipse periodicity.}
\label{fig:colorRVs}
\end{figure*}

\begin{figure*}
\centering
\includegraphics[width=1\textwidth]{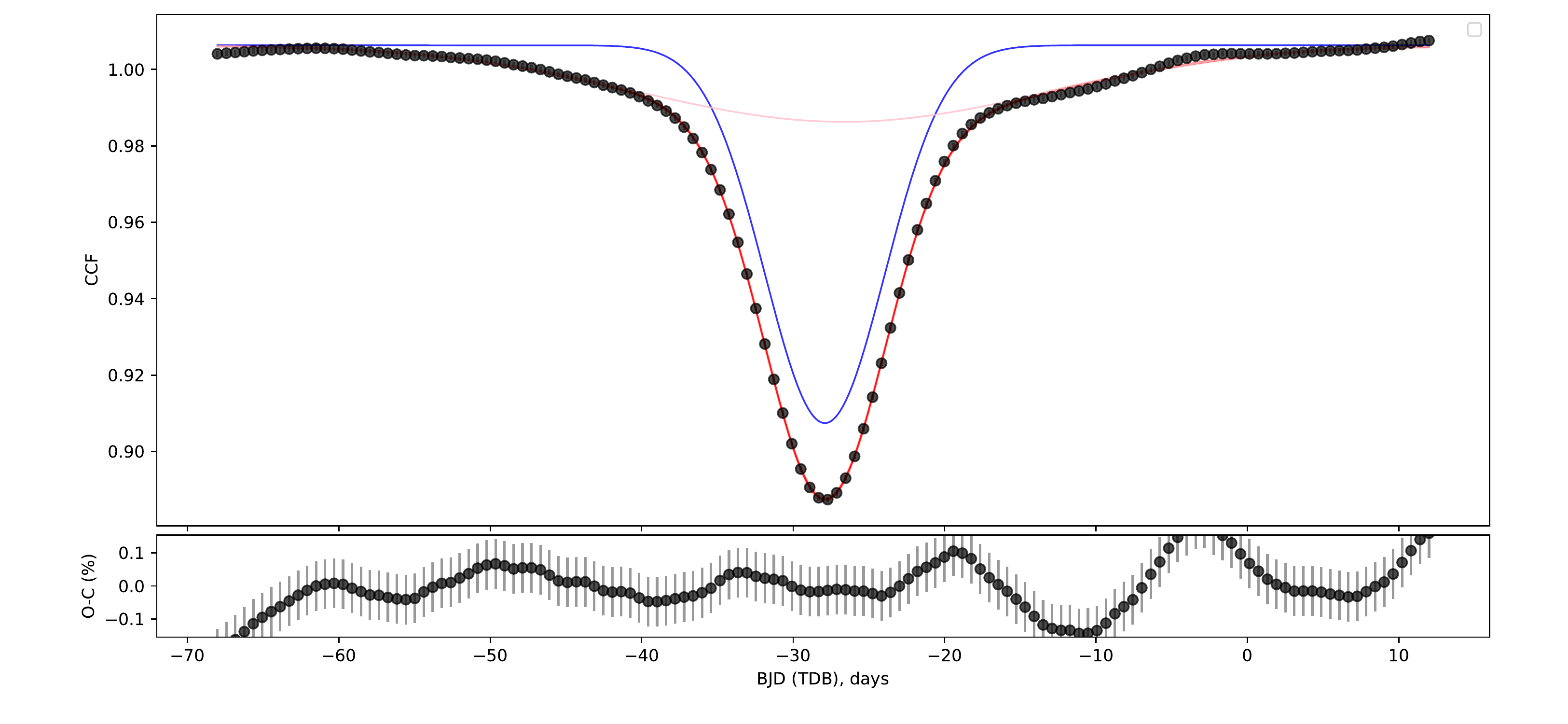}
\caption{Example of the CCF of one of the HERMES spectra (black symbols). The two-component modeling of the CCF is shown as a red line, while the two individual components are shown in blue (sharp component, star A) and light pink (broad component, star B). The bottom panel shows the residuals after subtracting the two-component model to the CCF.}
\label{fig:two-ccf}
\end{figure*}

\begin{figure}
\centering
\includegraphics[width=0.5\textwidth]{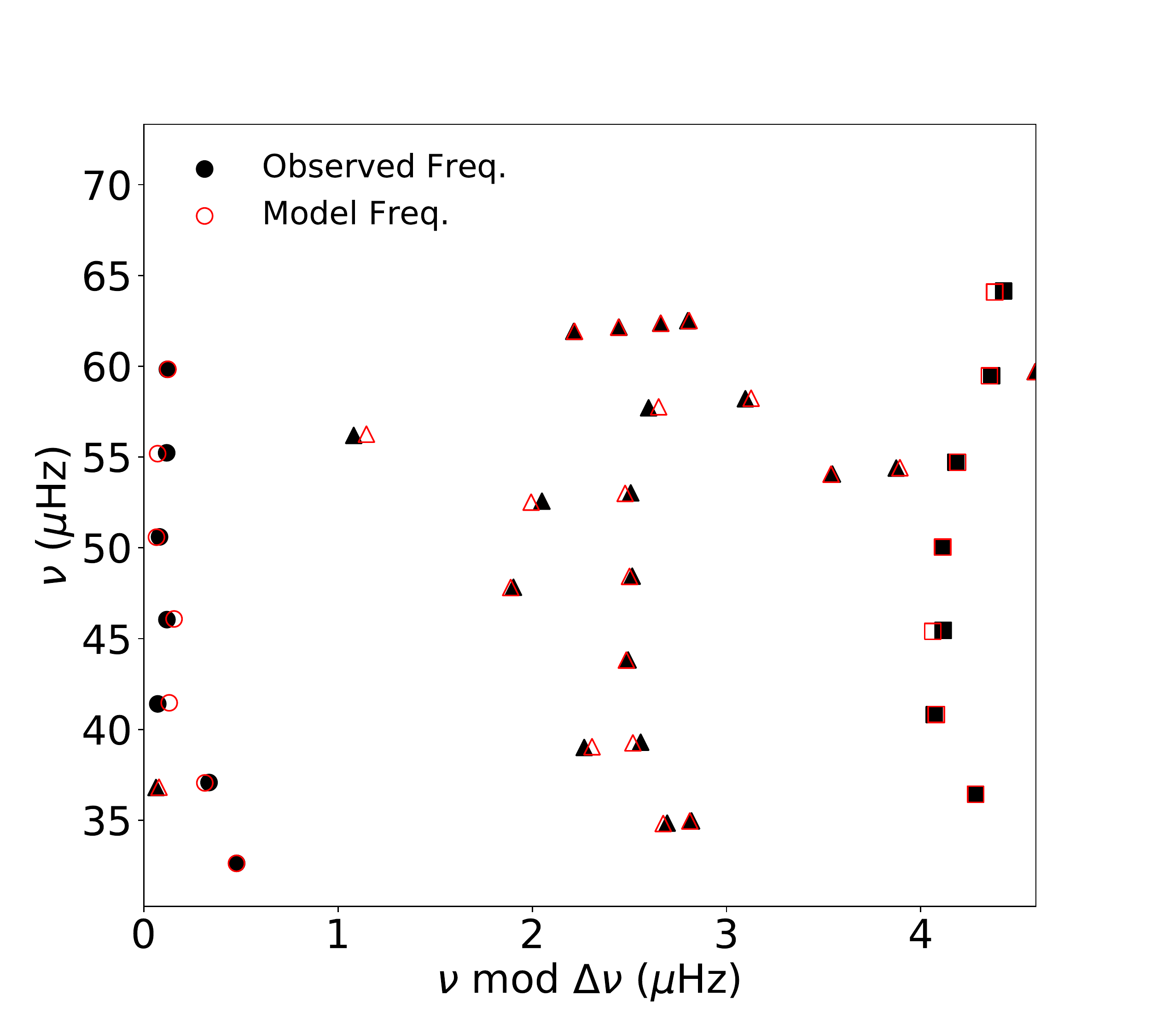}
\caption{\'Echelle diagram representing the observed frequencies (filled black symbols) as well as the model frequencies (open red symbols) corresponding to a representative best-fitting model. No interpolation was used and so this corresponds to a specific model in the grid. The circles, triangles, and squares indicate the modes of angular degree $\ell=0$ (radial modes), $\ell=1$ (dipole modes), and $\ell=2$ (quadrupole modes), respectively.}
\label{fig:echelle}
\end{figure}

\begin{figure*}
\centering
\includegraphics[width=1\textwidth]{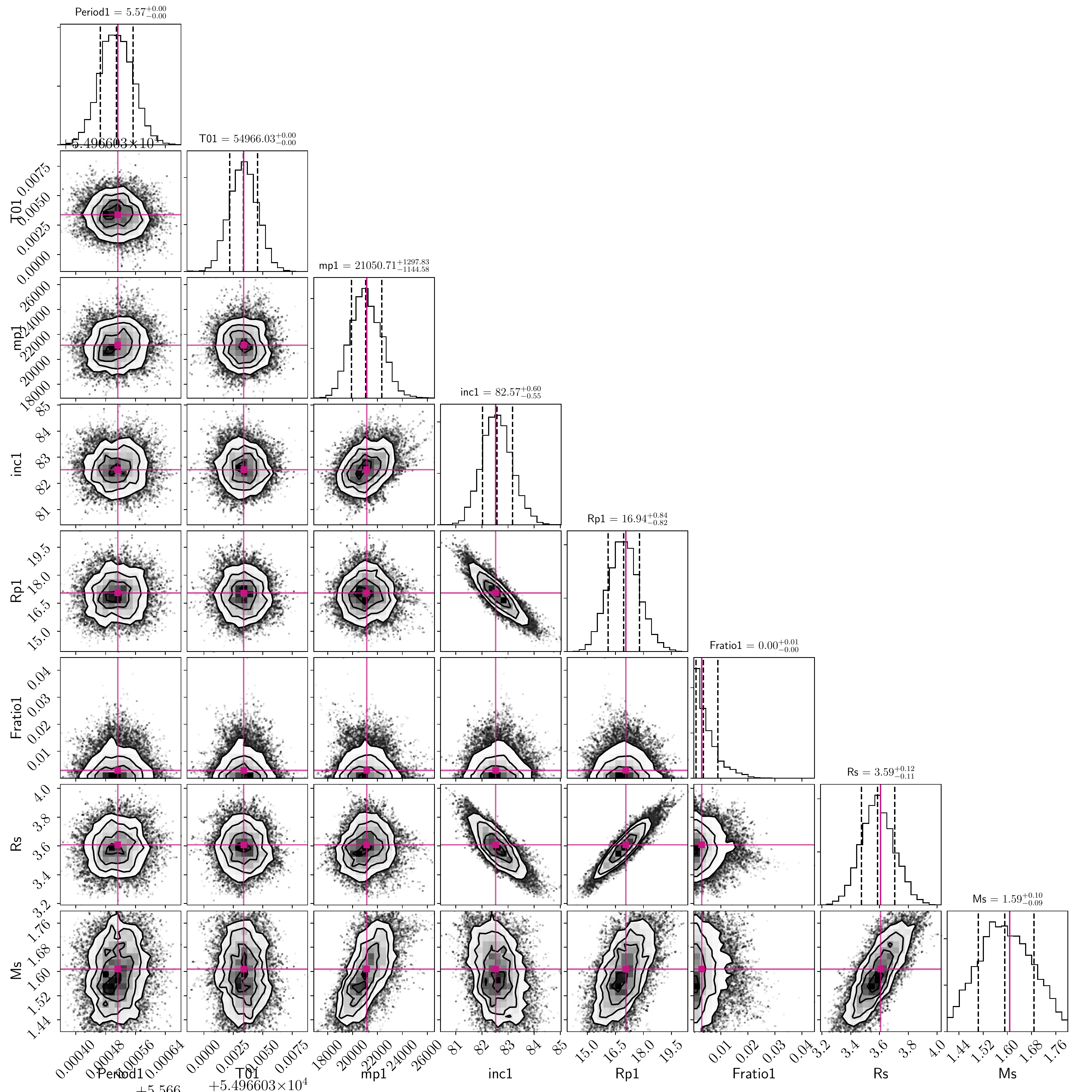}
\caption{Corner plot of the joint light curve and RV analysis of the KOI-3886 system.}
\label{fig:corner}
\end{figure*}

\onecolumn


\end{document}